\newcommand{\diag}{{\rm diag}}
\newcommand{\e}{{\rm e}}
\newcommand{\f}{f}
\newcommand{\third}{ {\textstyle{1\over 3}}}
\newcommand{\fourth}{ {\textstyle{1\over 4}}}
\newcommand{\twothirds}{ {\textstyle{2\over 3}}}
\newcommand{\uu}{u} 
\newcommand{\oms}{\om_0} 
\newcommand{\omp}{\om_{1/2}} 
\newcommand{\be}{\begin{equation}}
\newcommand{\ee}{\end{equation}}
\newcommand{\bea}{\begin{eqnarray}}
\newcommand{\eea}{\end{eqnarray}}
\newcommand{\ba}{\begin{array}}
\newcommand{\ea}{\end{array}}
\newcommand{\al}{\alpha}
\newcommand{\ga}{\gamma}
\newcommand{\de}{\delta}
\newcommand{\ep}{\epsilon}

\newcommand{\om}{\omega}

\newcommand{\la}{\lambda}

\newcommand{\vphi}{\varphi}

\newcommand{\hR}{\tilde{R}}
\newcommand{\hT}{\tilde{T}}

\newcommand{\hphi}{\tilde{\phi}}
\newcommand{\hPhi}{\tilde{\Phi}}
\newcommand{\hb}{\tilde{b}}
\newcommand{\hB}{\tilde{B}}
\newcommand{\hW}{\tilde{W}}
\newcommand{\hi}{\tilde{\imath}}
\newcommand{\hj}{\tilde{\jmath}}
\newcommand{\hk}{\tilde{k}}
\newcommand{\hl}{\tilde{l}}
\newcommand{\hm}{\tilde{m}}
\newcommand{\hr}{\tilde{r}}
\newcommand{\hs}{\tilde{s}}
\newcommand{\hp}{\tilde{p}}
\newcommand{\hq}{\tilde{q}}
\newcommand{\hn}{\tilde{n}}

\newcommand{\hla}{\tilde{\lambda}}
\newcommand{\hS}{\tilde{S}}
\newcommand{\hV}{\tilde{V}}
\newcommand{\hM}{\tilde{M}}
\newcommand{\hcW}{\tilde{\mathcal{W}}}
\newcommand{\home}{\tilde{\omega}}
\newcommand{\tx}{\tilde{x}}
\newcommand{\tX}{\tilde{X}}
\newcommand{\ttpi}{\frac{1}{32\pi^2}}
\def\gs{ g_{\rm s} }
\def\vev#1{ \langle {#1} \rangle }
\def\lvev#1{ \left\langle {#1} \right\rangle }
\def\bvev#1{ \bigg\langle {#1} \bigg\rangle }
\newcommand{\tr}{{\rm tr}}
\newcommand{\PP}{\mathrm{I}\kern -2.2pt \mathrm{P}}
\newcommand{\R}{\mathrm{I}\kern -2.2pt \mathrm{R}}
\newcommand{\sPP}{\mathrm{I}\kern -1.5pt \mathrm{P}}
\newcommand{\sR}{\mathrm{I}\kern -1.5pt \mathrm{R}}
\newcommand{\Z}{\mathsf{Z}\kern -5pt \mathsf{Z}}
\newcommand{\C}{\mathsf{I}\kern -5pt \mathrm{C}}
\newcommand{\D}{{\rm d}}
\newcommand{\pa}{\partial}
\newcommand{\rar}{\rightarrow}
\newcommand{\non}{\nonumber}

\newcommand{\cN}{\mathcal{N}}
\newcommand{\cW}{\mathcal{W}}

\newcommand{\cO}{\mathcal{O}}

\newcommand{\half}{{\textstyle {1\over 2}}}
\newcommand{\bD}{\bar{D}}

\newcommand{\1}{1\kern -3pt \mathrm{l}}
\newcommand{\SU}{\mathrm{SU}}
\newcommand{\SO}{\mathrm{SO}}
\newcommand{\Sp}{\mathrm{Sp}}

\newcommand{\U}{\mathrm{U}}

\documentclass[12pt]{article}

\usepackage{graphicx}

\newdimen\tableauside\tableauside=1.0ex
\newdimen\tableaurule\tableaurule=0.4pt
\newdimen\tableaustep
\def\phantomhrule#1{\hbox{\vbox to0pt{\hrule height\tableaurule
width#1\vss}}}
\def\phantomvrule#1{\vbox{\hbox to0pt{\vrule width\tableaurule
height#1\hss}}}
\def\sqr{\vbox{%
  \phantomhrule\tableaustep
\hbox{\phantomvrule\tableaustep\kern\tableaustep\phantomvrule\tableaustep}%
  \hbox{\vbox{\phantomhrule\tableauside}\kern-\tableaurule}}}
\def\squares#1{\hbox{\count0=#1\noindent\loop\sqr
  \advance\count0 by-1 \ifnum\count0>0\repeat}}
\def\tableau#1{\vcenter{\offinterlineskip
  \tableaustep=\tableauside\advance\tableaustep by-\tableaurule
  \kern\normallineskip\hbox
    {\kern\normallineskip\vbox
      {\gettableau#1 0 }%
     \kern\normallineskip\kern\tableaurule}%
  \kern\normallineskip\kern\tableaurule}}
\def\gettableau#1 {\ifnum#1=0\let\next=\null\else
  \squares{#1}\let\next=\gettableau\fi\next}
\newcommand{\Yfund}{\tableau{1}}
\newcommand{\Ysymm}{\tableau{2}}
\newcommand{\Yasymm}{\tableau{1 1}}
\newcommand{\sYasymm}{\tableauside=1.0ex \tableau{1 1}}

\begin{document}

\begin{flushright} 
{\tt hep-th/0303268}\\ 
BRX-TH-515 \\ 
BOW-PH-128\\
\end{flushright}
\vspace{1mm} 
\begin{center}
{\bf\Large\sf Cubic curves from matrix models
and generalized Konishi anomalies} 

\vskip 8mm 

Stephen G. Naculich\footnote{Research
supported in part by the NSF under grant PHY-0140281.}$^{,a}$,
Howard J. Schnitzer\footnote{Research
supported in part by the DOE under grant DE--FG02--92ER40706.}$^{,b}$,
and Niclas Wyllard\footnote{Research 
supported by the DOE under grant DE--FG02--92ER40706.\\
{\tt \phantom{aaa} naculich@bowdoin.edu;
schnitzer,wyllard@brandeis.edu}\\}$^{,b}$

\end{center}
\vspace{.05in}

\begin{center}
$^{a}${\em Department of Physics\\
Bowdoin College, Brunswick, ME 04011}

\vspace{.2in}

$^{b}${\em Martin Fisher School of Physics\\
Brandeis University, Waltham, MA 02454}

\end{center}
\vspace{.1in}

\begin{abstract} 
We study the matrix model/gauge theory connection for three different 
$\cN=1$ 
models: 
$\U(N){\times}\U(N)$ with matter in bifundamental representations,
$\U(N)$ with matter in the symmetric representation, and 
$\U(N)$ with matter in the antisymmetric representation. 
Using Ward identities, we explicitly
show that the loop equations of the matrix models lead 
to cubic algebraic curves.
We then establish the equivalence of the matrix model 
and gauge theory descriptions in two ways.
First, we derive generalized Konishi anomaly equations in the gauge theories,
showing that they are identical to the matrix-model equations. 
Second, we use a perturbative superspace analysis 
to establish the relation between the gauge theories and the matrix models. 
We find that the gauge coupling matrix 
for $\U(N)$ with matter in the symmetric or antisymmetric representations
is {\it not} given by the second derivative of the matrix-model free energy. 
However, 
the matrix-model prescription can be modified to 
give the gauge coupling matrix.
\end{abstract}

\setcounter{equation}{0}
\section{Introduction}
The matrix model approach \cite{Dijkgraaf:2002c} has provided a 
new way of studying (the holomorphic sector of) 
supersymmetric gauge theories. 
That the matrix model leads to results identical to
those of the gauge theory has been shown 
for the simplest model ($\U(N)$ with adjoint matter) using two methods. 
First, a remarkably succinct perturbative superspace argument was used 
to show \cite{Dijkgraaf:2002e} that the effective 
superpotential is equal to the corresponding matrix-model quantity 
order-by-order in a perturbative expansion in powers of the glueball field. 
Second, it was shown \cite{Cachazo:2002b}
that the (quadratic) loop equation of the matrix model is 
realized in the chiral ring of the gauge theory as a 
generalization of the Konishi 
anomaly equation~\cite{Konishi:1984}, thus 
establishing the (non-perturbative) 
correctness of the matrix-model description. 
The latter method was extended to include fundamental matter 
in ref.~\cite{Seiberg:2002}. 
The perturbative method can also be used to treat this case, 
although it was treated in less detail in ref.~\cite{Dijkgraaf:2002e}.
Some related earlier work and more recent developments 
can be found in 
refs.~\cite{Gorsky:2002,Cachazo:2003b}.

In this work, we extend the 
matrix model/gauge theory equivalence to three $\cN=1$ 
theories:\footnote{These theories have in common that, 
in the $\cN=2$ limit, they
all possess non-hyperelliptic (cubic) Seiberg-Witten curves.}
$\U(N){\times}\U(N)$ gauge theory with matter in adjoint 
and bifundamental representations,
$\U(N)$ gauge theory with matter in the 
adjoint and symmetric representations, and
$\U(N)$ gauge theory with matter in the 
adjoint and antisymmetric representations. 
We derive the cubic relations 
\be
u^3 - r(z) \, u - s(z) = 0 \,,
\ee
satisfied by the resolvents of the associated matrix models,
and give explicit expressions for the coefficients of the 
polynomials $r(z)$ and $s(z)$ 
in terms of the adjoint-field eigenvalues, 
using a Ward-identity approach.
These loop equations encode the geometry of cubic algebraic curves 
underlying these models. On 
the gauge theory side we consider generalized Konishi anomaly equations 
and show that they lead to equations identical 
to the matrix-model loop equations, thus establishing the equivalence.   
We also use a perturbative superspace analysis to analyze  
the relation between the 
gauge theories and the matrix models. 
We find that for the $\U(N)$ models with matter in the 
symmetric/antisymmetric representations, the gauge coupling matrix is 
{\it not}
given by the second derivative of the matrix-model free energy. 
Nevertheless, 
the matrix-model prescription can be modified to give
the gauge theory coupling matrix.

Various aspects of the $\U(N){\times}\U(N)$ model were discussed 
in refs.~\cite{Dijkgraaf:2002b,Tachikawa:2002,Hofman:2002}
and also recently in ref.~\cite{Lazaroiu:2003}.
The $\U(N)$ models with symmetric or antisymmetric matter 
were also studied recently in ref.~\cite{Klemm:2003}. 
There is some overlap between the present work and 
the recent papers~\cite{Lazaroiu:2003, Klemm:2003}, but for the most part 
our work is complementary to their analysis. The 
explicit expressions for $s(z)$ that we derive in this paper
were also obtained in \cite{Lazaroiu:2003,Klemm:2003} 
(using a different method);    
however, the gauge theory analogs of the loop equations were 
not discussed and the equivalence was not established.

In sec.~\ref{sU2},
we discuss the supersymmetric $\U(N){\times}\U(N)$ theory with 
bifundamental matter. In sec.~\ref{sUSAI},
we perform a similar analysis for the 
supersymmetric $\U(N)$ gauge theory with matter in 
symmetric or in antisymmetric representations. 
In sec.~\ref{sUSAII}, we use superspace perturbation 
theory to analyze the $\U(N)$ models. 
A summary of the main results of the paper can be found in sec.~\ref{sSum}.
In the appendices, we briefly discuss the 
saddle-point approach as an alternative to
the method used in the main text,
and collect some background material on the relevant representations. 

\setcounter{equation}{0}
\section{{\large $\U(N){\times}\U(N)$} with bifundamental matter} 
\label{sU2}

In this section we study the $\cN=1$ $\U(N){\times}\U(N)$ 
supersymmetric gauge theory with the 
following matter content: 
two chiral superfields 
$\phi_i{}^j$, $\hphi_{\hi}{}^{\hj}$ transforming 
in
the adjoint representation of each of the two factors of the gauge group, 
one chiral superfield $b_i{}^{\hj}$ 
transforming in the bifundamental representation $(\Yfund,\bar{\Yfund})$, and 
one chiral superfield $\hb_{\hi}{}^j$ transforming in the 
bifundamental representation $(\bar{\Yfund},\Yfund)$.
The superpotential of the gauge theory is taken to be of the 
form\footnote{\label{massfoot}An explicit mass term 
for the bifundamental field, 
$ m \, \tr( b \hb)$, can be introduced by shifting $\phi$ and 
$\hphi$ and redefining 
the coefficients in $W(\phi)$ and $\hW(\hphi)$; to simplify the 
presentation we will therefore not 
explicitly include such a term, although we think of the bifundamental 
fields as being massive.}

\be \label{U2suppot}
\cW(\phi,\hphi,b,\hb) = \tr \,[ W(\phi) - \hW(\hphi) - \hb \, \phi \,b 
+ b\, \hphi \,\hb ] \,,
\ee
where $W(\phi) = \sum_{m=1}^{N+1} (g_m/m) \, \phi^m$ 
and similarly for $\hW(\hphi)$. 
This superpotential can be viewed as a deformation of an $\cN=2$ theory.

Below, after a detailed derivation of the loop equations of the matrix model,
we establish the non-perturbative equivalence between the 
holomorphic sector of  the above gauge theory and the associated matrix model, 
following the ideas developed in ref.~\cite{Cachazo:2002b}. 
(The argument for the perturbative equivalence of the matrix model
and gauge theory given in~\cite{Dijkgraaf:2002e} goes through 
essentially unchanged for this case.)
More precisely, we show that the matrix-model loop equations 
are encoded in the gauge theory as vacuum expectation values 
of divergences of certain anomalous currents. 
The anomalies associated with these currents are 
generalizations of the Konishi anomaly \cite{Konishi:1984}.

\subsection{Matrix model analysis}

Following the ideas of Dijkgraaf and Vafa, we take the partition function for 
the matrix model associated with the above gauge theory to 
be\footnote{We use capital letters to denote matrix model quantities.}
\be 
\label{U2Z}
Z = \int \D \Phi \, \D \hPhi \, \D B \, \D \hB \, 
\exp \!\left( -\frac{1}{\gs} 
\tr\!\left[W(\Phi) - \hW(\hPhi) - \hB \Phi B + B \hPhi \hB \right] \right)\,,
\ee
where $\Phi$ is an $M{\times}M$ matrix, $\hPhi$ is an $\hM{\times}\hM$ matrix, 
$B$ is an $M{\times}\hM$ matrix, and $\hB$ is an $\hM{\times}M$ matrix.
These matrices should be viewed as holomorphic 
quantities~\cite{Dijkgraaf:2002a,Cachazo:2002b,Lazaroiu:2003} 
and the integrals in (\ref{U2Z}) are along some curve. 
This point was emphasized in the recent 
paper \cite{Lazaroiu:2003}, where 
the above model was also studied.  
We are interested in the planar limit of the matrix model, i.e.~the 
limit in which $\gs \rar 0$ and $M$, $\hM$ $\rar \infty$, keeping 
$S=\gs M$ and $\hS = \gs \hM$ fixed. 

In the saddle-point approach to this model
\cite{Kostov:1992,Kharchev:1993,Dijkgraaf:2002b,
Hofman:2002,Lazaroiu:2003},
one diagonalizes the matrices $\Phi$ and $\hPhi$,
and derives equations satisfied by the resolvents\footnote{We use an
unconventional normalization of the resolvents in order to 
make the comparison with gauge theory more transparent.
Also, in order 
not to clutter the formul\ae{} we drop the $\vev{\cdots}$ when writing 
expressions in terms of eigenvalues.} 
\be \label{ress}
\om(z) = \gs \bvev{ \tr\!\left(\frac{1}{z-\Phi}\right) } = 
\gs \sum_i \frac{1}{z-\la_i} \, ; \quad 
\home(z) = \gs \bvev{ \tr\!\left(\frac{1}{z-\hPhi}\right) } = 
\gs \sum_i \frac{1}{z-\hla_i} \,,
\ee
where matrix-model expectation values are defined via
\be \label{U2MMexp}
\bvev{\cO(\Phi,\hPhi,B,\hB)} 
\equiv
\frac{1}{Z} \int \D \Phi \, \D \hPhi \,  \D B \, \D \hB \, 
\cO(\Phi,\hPhi,B,\hB) \, 
\e^{  -\frac{1}{\gs}
\tr\left[W(\Phi) - \hW(\hPhi) - \hB \Phi B + B \hPhi \hB \right] }\,.
\ee
For completeness we give some 
details of the saddle-point approach in appendix \ref{appcon}. 

Below we derive the equations satisfied by the resolvents
using an approach \cite{Cachazo:2002b}
that is close in spirit to the gauge theory analysis 
given in section \ref{sgU2}. 
(We stress that this method does not assume that the matrices are hermitian.)

Throughout the paper we often suppress matrix indices, assuming that 
multiplications are done using the natural contractions. 

\medskip
\noindent {\bf Quadratic relations}
\medskip 

We start by considering the Ward identity
\bea \label{homeom}
&& \!\! 0\, 
= \, \frac{\gs^2}{Z} \int \D \Phi \, \D \hPhi \,  \D B \, \D \hB \, 
\frac{\D}{\D B_i{}^{\hj}} 
\left\{  
\left(\frac{1}{z-\Phi}B\frac{1}{z-\hPhi}
\right)_{\raisebox{5pt}{\scriptsize $i$}}
{}^{\raisebox{2pt}{\scriptsize $\hj$}} 
\, \e^{ -\frac{1}{\gs}\tr[W(\Phi) - \hW(\hPhi) 
- \hB \Phi B + B \hPhi \hB ] } \right\}  \non \\
&& \!\! = \gs^2 
\bvev{ \! \tr \! \left(\frac{1}{z-\Phi}\right)
\tr \! \left(\frac{1}{z-\hPhi}\right) \! }
+ \gs \bvev{ \! \tr \! \left( \frac{\Phi}{z-\Phi} B 
\frac{1}{z-\hPhi} \hB \right) \! }
- \gs \bvev{ \! \tr \! \left(\frac{1}{z-\Phi} 
B \frac{\hPhi}{z-\hPhi}\hB \right) \! } \non \\
&& \!\! = \om(z)\, \home(z) - \gs \bvev{ \! \tr \! 
\left( B \frac{1}{z-\hPhi}\hB \right) \! } 
+ \gs \bvev{ \!\tr\!\left( \hB \frac{1}{z-\Phi} B \right)\!} \,,
\eea  
where the resolvents were defined in (\ref{ress}), and we have used 
(here and throughout) the factorization of expectation values 
in the planar limit. Thus, the expectation values and resolvents 
appearing in (\ref{homeom}) (and in all remaining equations 
in this section) refer only to the planar (sphere) parts in the 
genus expansion; we will not indicate this explicitly as confusion 
is unlikely to arise.

Next, we note that for any polynomial $f(z)$, we have
the Ward identity
\bea \label{homerphi}
&& \!\!\!\! 0 
\,=\, \frac{\gs^2}{Z} \int \D \Phi \, \D \hPhi \,  \D B \, \D \hB \, 
 \frac{\D}{\D \Phi_i{}^j} 
\left\{ \left(\frac{\f(\Phi)}{z-\Phi}
\right)_{\raisebox{5pt}{\scriptsize $i$}}
{}^{\raisebox{2pt}{\scriptsize $j$}}  \, 
\e^{ -\frac{1}{\gs}\tr[W(\Phi) - \hW(\hPhi) 
- \hB \Phi B + B \hPhi \hB ] }  \right\} \\
&& \!\!\!\!= 
  \om(z)^2 \f(z) 
- \gs^2 \bvev { \!\tr\! \left[ \frac{\D}{\D \Phi} 
\left(\frac{\f(z) - \f(\Phi)}{z-\Phi}\right) \right] \!} 
- \gs \bvev{\!\tr\!\left(\frac{\f(\Phi) W'(\Phi)}{z-\Phi}\right) \!}
+ \gs \bvev{\! \tr\!\left( \hB  \frac{\f(\Phi)}{z-\Phi} B \right)\!}\non .
\eea
In particular, setting $f(\Phi)=1$, eq.~(\ref{homerphi}) simplifies to 
\be \label{exp2}
\om(z)^2  - W'(z) \, \om(z) = 
- \gs \bvev{\tr\!\left(\frac{W'(z) - W'(\Phi)}{z-\Phi}\right) }
- \gs \bvev{ \tr\!\left(\hB  \frac{1}{z-\Phi} B \right)}\,.
\ee
For future reference we also note that by 
multiplying eq.~(\ref{exp2}) by $\f(z)$
and combining the resulting expression with eq.~(\ref{homerphi}) one finds
\be\label{later1}
\gs \bvev{ \!\tr\!\left( \hB  \frac{\f(z)-\f(\Phi)}{z-\Phi} B \right)\!}
=
- \gs^2 \bvev { \!\tr\! \left[ \frac{\D}{\D \Phi} 
\left(\frac{\f(z) - \f(\Phi)}{z-\Phi}\right) \right] \!} 
+ \gs \bvev{\!\tr\!\left(\frac{\f(z)-\f(\Phi)}{z-\Phi} W'(\Phi)\right)\! }.
\ee

Analogously, one can show
\be \label{omrhphi}
0 = 
  \home(z)^2 \f(z) 
-\gs^2 \bvev{\!\tr\!\left[\frac{\D}{\D \hPhi} 
\left(\frac{\f(z) - \f(\hPhi)}{z-\hPhi}\right)\right] \!} 
+ \gs \bvev{\!\tr\!\left(\frac{\f(\hPhi) \hW'(\hPhi)}{z-\hPhi}\right)\! }
- \gs \bvev{ \!\tr\!\left(B  \frac{\f(\hPhi)}{z-\hPhi} \hB \right)\!},
\ee
{}from which it follows by setting $f(\hPhi)=1$ that
\be \label{exp3}
\home(z)^2 + \hW'(z)\, \home(z) 
=  \gs \bvev{\!\tr\!\left(\frac{\hW'(z)-\hW'(\hPhi)}{z-\hPhi}\right)\! }  
+ \gs \bvev{\!\tr\!\left(B \frac{1}{z-\hPhi} \hB\right)\!}\,.
\ee
By combining the previous two equations, we obtain
\be
\label{later2}
\gs \bvev{ \!\tr\!\left( B  \frac{\f(z)-\f(\hPhi)}{z-\hPhi} \hB  \right)\!}
= \gs^2 \bvev { \!\tr\! \left[ \frac{\D}{\D \hPhi} 
\left(\frac{\f(z) - \f(\hPhi)}{z-\hPhi}\right) \right] \!} 
+ \gs 
\bvev{\!\tr\!\left(\frac{\f(z)-\f(\hPhi)}{z-\hPhi} \hW'(\hPhi)\right) \!}.
\ee

{}From the above equations it is possible to derive a quadratic relation
among the resolvents that does not involve expectation values 
with $B$, $\hB$'s.  
Combining eqs.~(\ref{homeom}), (\ref{exp2}), and (\ref{exp3}) 
to eliminate the $B$-dependent terms,
one obtains the following quadratic relation involving the two resolvents
\be \label{U2r1eq}
\om(z)^2 + \home(z)^2 - \om(z)\, \home(z) 
- W'(z) \, \om(z) + \hW'(z) \, \home(z) 
= r_1(z) \,,
\ee
where
\bea \label{U2r1}
r_1(z) &=&  - \gs \bvev{\tr\!\left(\frac{W'(z)-W'(\Phi)}{z-\Phi}\right) } 
+ \gs \bvev{\tr\!\left(\frac{\hW'(z)-\hW'(\hPhi)}{z-\hPhi}\right) }  \non\\
&=& -  \gs \sum_{i}
\frac{W'(z) - W'(\la_i)}{z-\la_i} 
 +   \gs  \sum_{i} \frac{\hW'(z) - \hW'(\hla_i)}{z-\hla_i} \,,
\eea
is a polynomial of degree at most $N-1$.

\medskip
\noindent{\bf The cubic algebraic curve}
\medskip

We now discuss how the cubic algebraic curve that underlies the 
model \cite{Dijkgraaf:2002b,Hofman:2002} emerges. One can eliminate 
the terms linear in the resolvents in eq.~(\ref{U2r1eq}) by defining
\be\label{U2omrelate}
\om(z) = \uu_1(z) + \om_r(z)\,, \qquad \qquad \home(z) = -\uu_3(z) 
+ \home_r(z)\,,
\ee
where
\be \label{U2omdef}
\om_r(z) = \twothirds W'(z)- \third \hW'(z) \,, \qquad\qquad
\home_r(z) =  \third W'(z) -\twothirds \hW'(z) \,,
\ee
giving 
\be \label{U2r0eq}
\uu_1(z)^2 + \uu_3(z)^2 + \uu_1(z) \uu_3(z) 
= r_0(z) +  r_1(z) = r(z) \,,
\ee
with
\bea\label{U2r0}
r_0  (z) 
&=& \om_r^2(z) + \home_r^2(z) - \om_r(z) \home_r(z) \non\\
&=& \third \left[ {W'}^{2}(z) + {\hW'}{}^2(z) - W'(z)\hW'(z) \right] \,,
\eea
a polynomial of degree $2N$.

Multiplying eq.~(\ref{U2r0eq}) by $\uu_1(z) - \uu_3(z)$, one finds
\be
\uu_1(z)^3 - r(z) \, \uu_1(z) = \uu_3(z)^3 - r(z) \,\uu_3(z) \equiv s(z)\,,
\ee
so that $\uu_1(z)$ and $\uu_3(z)$ are both roots of the cubic equation
\be\label{U2cubic}
0 = u^3 - r(z) \, u - s(z)  = [u-\uu_1(z)] [u-\uu_2(z)] [u-\uu_3(z)]\,. 
\ee
The absence of the quadratic term implies that the third root
is $\uu_2(z) = - \uu_1(z) - \uu_3(z)$, so
\be\label{U2sdef}
s(z) = \uu_1(z) \, \uu_2(z) \, \uu_3(z) 
= [ \om(z) - \om_r(z) ] [ -\om(z) +\home(z) + \om_r(z) - \home_r(z)]
[-\home(z) + \home_r(z)],
\ee
which we will show to be a polynomial below.

Defining $s(z)  = s_0(z) + s_1(z)$ with 
\bea\label{U2s0def}
s_0(z) &=& - \om_r(z) \,\home_r(z) \,[\om_r(z) - \home_r(z)]  \non\\
      &=& {\textstyle{1 \over 27}} 
       [- W'(z)+2\hW'(z)][2W'(z) - \hW'(z)] [W'(z) + \hW'(z)]\,,
\eea
a polynomial of degree $3N$,
we can rewrite the cubic equation (\ref{U2cubic}) as
\bea
r_1(z) u +  s_1(z)
&=& u^3 - r_0(z) u - s_0(z)  \non\\
&=& (u+\om_r(z)) (u-\om_r(z)+\home_r(z)) (u-\home_r(z)) \,.
\eea
{}From eqs.~(\ref{U2sdef}) and (\ref{U2s0def}) it follows that
\bea \label{U2s1eq}
s_1(z) &=& \, \om(z) \, \home(z) [\om(z) - \home(z)] 
-\twothirds  [ W'(z) + \hW'(z)] \om(z)\, \home(z)   \non\\
&& -\,\home_r(z)\,  [\om(z)^2 - W'(z) \om(z) ] 
+ \om_r(z) \, [\home(z)^2 + \hW'(z) \home(z)]\,.
\eea
We will show below that $s_1(z)$ is a polynomial of degree at most $2N{-}1$.

\medskip
\noindent{\bf Cubic relations}
\medskip

Above we studied Ward identities leading to expressions with at 
most two bifundamental fields. 
We will now analyze expressions involving 
two additional bifundamental fields. The resulting equations can be used 
to derive a cubic relation among the resolvents of the 
form (\ref{U2s1eq}). 
The fact that one need not
consider Ward identities with an even larger number of  
bifundamental fields can be traced to the form of the 
potential (\ref{U2suppot}).

Our starting point is the Ward identity
\bea \label{stepI}
0 &=& \frac{\gs^2}{Z} \int \D \Phi \, \D \hPhi \,  \D B \, \D \hB \, 
 \frac{\D}{\D \hPhi_{\hi}{}^{\hj}}  
\left\{   \left( \hB B \frac{1}{z-\hPhi}\right)
_{\raisebox{5pt}{\scriptsize $\hi$}}{}^{\raisebox{2pt}{\scriptsize $\hj$}}
\, \e^{ -\frac{1}{\gs}\tr[W(\Phi) - \hW(\hPhi) 
- \hB \Phi B + B \hPhi \hB ] }  \right\} \\
&=& \gs \, \home(z) \bvev{ \tr\!\left(B \frac{1}{z-\hPhi} \hB \right)} 
+ \gs \bvev{ \tr\!\left(B \frac{\hW'(\hPhi)}{z-\hPhi} \hB\right) } 
- \gs \bvev{\tr\!\left( \hB B \frac{1}{z-\hPhi} \hB B\right)}\non \,.
\eea
Similarly,
\bea \label{stepII}
0 = \gs \, \om(z) \bvev{ \tr\!\left(\hB \frac{1}{z-\Phi}B \right)} 
- \gs \bvev{ \tr\!\left(\hB \frac{W'(\Phi)}{z-\Phi} B\right) } 
+ \gs \bvev{ \tr\!\left(B \hB \frac{1}{z-\Phi} \hB B\right)} \,.
\eea
We will also need
\bea
\label{stepIII}
0 &=& \frac{\gs^2}{Z} \int \D \Phi \, \D \hPhi \,  \D B \, \D \hB \, 
\frac{\D}{\D \hB_{\hi}{}^{j}}  \left\{
\left( \frac{1}{z-\hPhi} \hB B \hB \frac{1}{z-\Phi} 
\right)_{\raisebox{5pt}{\scriptsize $\hi$}}
{}^{\raisebox{2pt}{\scriptsize $j$}} 
\, 
\e^{ -\frac{1}{\gs}\tr[W(\Phi) - \hW(\hPhi) 
- \hB \Phi B + B \hPhi \hB ] } \right\}  \non\\
&=& \gs \, \home(z) \bvev{\tr\!\left(\hB \frac{1}{z-\Phi} B\right) }
+  \gs \, \om(z) \bvev{ \tr\!\left(B \frac{1}{z-\hPhi} \hB \right)} \non \\
&&- \gs \bvev{\tr\!\left(\hB B \frac{1}{z-\hPhi}\hB B\right)} 
+ \gs \bvev{ \tr\!\left(B \hB \frac{1}{z-\Phi} B \hB \right)} \,. 
\eea

One can eliminate the terms quartic in the bifundamental fields from 
the above three equations; by combining (\ref{stepI}), (\ref{stepII}), 
and (\ref{stepIII}) and also using  eq.~(\ref{homeom}), one finds
\be \label{s1one}
\om(z)\, \home(z) \left[\om(z) - \home(z) \right]
= - \gs \bvev{\tr\!\left( \hB \frac{W'(\Phi)}{z-\Phi} B \right)}
+ \gs \bvev{ \tr\!\left(B \frac{\hW'(\hPhi)}{z-\hPhi}\hB\right) } \,. 
\ee
Finally, by using eqs.~(\ref{homeom}), (\ref{exp2}), (\ref{later1}), 
(\ref{exp3}), and (\ref{later2})
to eliminate the remaining 
dependence on the bifundamental fields, 
one obtains the cubic relation (\ref{U2s1eq}) with the following explicit 
expression for 
$s_1(z)$ 
\bea  \label{U2s1Phi}
s_1(z) &=& 
\gs\,\home_r(z) \bvev{\tr\!\left(\frac{W'(z)-W'(\Phi)}{z-\Phi}\right)}
+\gs \,\om_r(z)
\bvev{\tr\!\left(\frac{\hW'(z)-\hW'(\hPhi)}{z-\hPhi}\right) }
\non\\
&& - \,\gs^2 \bvev { \tr\!\left[ \frac{\D}{\D \Phi}
\left(\frac{W'(z) - W'(\Phi)}{z-\Phi}\right) \right] } 
  - \gs^2 \bvev {\tr\!\left[\frac{\D}{\D \hPhi} 
\left(\frac{\hW'(z) - \hW'(\hPhi)}{z-\hPhi}\right)\right] }  \non\\
&&
+ \,\gs \bvev{\tr\!\left(\frac{W'(z)-W'(\Phi)}{z-\Phi} W'(\Phi)\right) }
- \gs \bvev{\tr\!\left(\frac{\hW'(z)-\hW'(\hPhi)}{z-\hPhi} \hW'(\hPhi)
\right) }.\,
\eea
At this point, it is clear that $s_1(z)$ is a polynomial
of degree at most $2N-1$, whose coefficients depend on the vevs 
$\vev{\tr(\Phi^k)}$ and $\vev{\tr(\hPhi^k)}$ 
with $k\le 2N{-}1$.

We will now write $s_1(z)$ more explicitly, in terms
of the eigenvalues $\la_i$ and $\hla_i$ of $\Phi$ and $\hPhi$ respectively.
First observe that, since 
$\frac{\f(z)-\f(\Phi)}{z-\Phi}  \equiv \sum_{m} c_m \Phi^m$ 
is a polynomial,
we have
\bea
\label{eigenvaluetrick}
&&\tr\left[ \frac{\D}{\D \Phi} 
\left(\frac{\f(z)-\f(\Phi)}{z-\Phi}\right)\right] 
= \sum_m c_m \sum_{k=0}^{m-1} \tr(\Phi^k) \tr(\Phi^{m-k-1}) 
= \sum_{i,j} \sum_m c_m \sum_{k=0}^{m-1} \la_i^k \la_j^{m-k-1} \non\\
&&\qquad = \sum_{i,j} \sum_m c_m \frac{\la_i^m-\la_j^m}{\la_i-\la_j} 
= \sum_{i,j} \frac{1}{\la_i-\la_j} \left[ \frac{\f(z)-\f(\la_i)}{z-\la_i} 
- \frac{\f(z)-\f(\la_j)}{z-\la_j} \right] 
\non\\&&\qquad 
= 2 \sum_{i \neq j} \frac{1}{\la_i-\la_j} 
\left[ \frac{\f(z)-\f(\la_i)}{z-\la_i} \right] \,.
\eea
Hence, we may write (suppressing $\vev{\cdots}$ in the eigenvalue basis)
\bea \label{U2s1}
s_1(z) &=&  \gs\, \home_r(z) \sum_i 
\left[ \frac{ W'(z)  - W'(\la_i)}{z-\la_i} \right]
+ \gs \,\om_r(z) \sum_i 
\left[ \frac{ \hW'(z)- \hW'(\hla_i)}{z-\hla_i}\right]
 \non \\
& - & 2\gs^2 \sum_{i \neq j}\frac{1}{\la_i-\la_j} 
\left[ \frac{W'(z) - W'(\la_i)}{z-\la_i} \right] 
 -  2  \gs^2  \sum_{i \neq j} \frac{1}{\hla_i-\hla_j}
\left[ \frac{\hW'(z) - \hW'(\hla_i)}{z-\hla_i} \right] 
\non\\
& + & \gs \sum_{i} W'(\la_i)
\left[ \frac{W'(z) - W'(\la_i)}{z-\la_i} \right] 
 -   \gs  \sum_{i}\hW'(\hla_i)
\left[ \frac{\hW'(z) - \hW'(\hla_i)}{z-\hla_i} \right] \,.
\eea
Finally, using the saddle point equations (\ref{U2saddle}), 
we may rewrite this as 
\bea \label{U2s1final}
s_1(z) &=&  \gs \, \home_r(z) \sum_i 
\left[\frac{ W'(z)  - W'(\la_i)}{z-\la_i}\right]
+ \gs \, \om_r(z) \sum_i 
\left[\frac{ \hW'(z)- \hW'(\hla_i)}{z-\hla_i} \right]
\non \\
&  - &  \gs^2 \sum_{i,j} \frac{1}{\la_i-\hla_j}
\left[\frac{W'(z)-W'(\la_i)}{z-\la_i} 
- \frac{\hW'(z)-\hW'(\hla_j)}{z-\hla_j} \right]\,, 
\eea
which is a polynomial of degree no more that $2N{-}1$.
This result also appeared recently in ref.~\cite{Lazaroiu:2003}, 
although using a different method 
of derivation.

This concludes our discussion of the $\U(N){\times} \U(N)$ matrix model. 
We now turn to the gauge theory analysis.

\subsection{Gauge theory analysis} \label{sgU2}

As will now be shown,  
the matrix-model loop equations can be obtained 
from certain generalizations of the Konishi anomaly equations 
in the gauge theory.
We find a one-to-one correspondence with the matrix-model 
formul\ae{} derived above.

It is sufficient to study the chiral part of the 
anomaly equations \cite{Cachazo:2002b}, i.e.~one may use 
identities that hold in the chiral ring. The chiral ring is defined as 
all chiral operators 
modulo terms of the form $\{ \bar{Q}_{\dot{\al}}, \cdot \}$.
For a Grassmann even field $F$,
one therefore has, in the chiral ring,
$0= [ \bar{Q}^{\dot{\al}},D_{\al\dot{\al}}F] = 
{\sf W}_\al F$, where ${\sf W}_\al$ is the (spinor) gauge superfield. 
As discussed in appendix \ref{apprep}, ${\sf W}_\al$ 
can be viewed as a diagonal $2{\times}2$ matrix 
where the entries along the diagonal are the gauge superfields of 
the two $\U(N)$ factors, $\cW_\al$ and $\hcW_\al$. 
More explicitly, ${\sf W}_\al = {\sf W}_{\al}^A T^A$, 
where $T^A$ are the representation matrices appropriate for the 
action of the gauge field on $F$. Using the explicit expressions for $T^A$ 
given in appendix \ref{apprep} one obtains identities (in the chiral ring) 
among the adjoint and bifundamental fields of the form (as anticipated 
in \cite{Cachazo:2002b})
\be \label{U2chirels}
{}[\cW_{\al},\phi]=0 \,, \qquad [\hcW_{\al},\hphi]=0
\,,\qquad \cW_{\al}\, b = b \,\hcW_{\al} \,, \qquad 
\hcW_{\al} \,\hb = \hb \,\cW_{\al}\,.
\ee
These identities will be freely used in what follows.
The Grassmann oddness of $\cW_\al$ 
together with the relations
$\cW_\al = \ep_{\al\beta}\cW^\beta$ and 
$\ep^{\beta \al}\ep_{\al\ga} = \de_\ga^\beta$ 
will also be used below.

The basic building blocks that we use are the anomalous currents
$\bar{\hb}_{\hi}{}^j (e^{\sf V} b)_k{}^{\hl}$, 
$\bar{\phi}_i{}^j (e^{\sf V}\phi)_k{}^l$,  
and $\hphi_{\hi}{}^{\hj} (e^{\sf V} \hphi)_{\hk}{}^{\hl}$,
where ${\sf V}$ is the (vector) gauge superfield 
(see appendix \ref{apprep} for more details about the notation).

We are interested in the action of $\bD^2$ on the currents. Using 
the superpotential (\ref{U2suppot}),
one finds the classical piece of 
$ \bD^2 \bar{\hb}_{\hi}{}^j (e^{\sf V} b)_k{}^{\hl} $ to be
$ [-(\hb \phi)_{\hi}{}^j + (\hphi \hb)_{\hi}{}^j] b_k{}^{\hl} $.
This current also has an anomaly (see 
appendix \ref{apprep} for an explanation of the notation)
\bea \label{bhbanom}
&& \frac{1}{32\pi^2} ({\sf W}_\al)_N{}^M ({\sf W}^\al)_Q{}^P 
(T_M{}^N)_{\hi}{}^j {}_m {}^{\hn} (T_P{}^Q)_{\hn} {}^m {}_k {}^{\hl} \non \\
&=& \frac{1}{32\pi^2}
\left[ (\hcW_\al \hcW^\al)_{\hi}{}^{\hl} \de_k^j 
- (\cW_\al)_k{}^j (\hcW^\al)_{\hi} {}^{\hl} 
- (\hcW_\al)_{\hi} {}^{\hl} (\cW^{\al})_k {}^j 
+ (\cW_\al \cW^\al)_k{}^j \de_{\hi}^{\hl} \, \right]
\eea
There might be perturbative corrections to the anomaly but these 
will be non-chiral \cite{Cachazo:2002b} and so are 
not of interest to us. We assume that there are no 
non-perturbative corrections to the anomaly.

The classical piece of $\bD^2 \bar{\phi}_i {}^j (e^{\sf V} \phi)_k {}^l $ 
is $ [W'(\phi)_i {}^j  - (b \hb)_i {}^j] \phi_{k}{}^{l} $
and the anomaly is 
\bea \label{phianom}
&& \frac{1}{32\pi^2} (\cW_\al)_s{}^r ( \cW^\al)_q{}^p 
(T_r{}^s)_{i}{}^j {}_m {}^{n} (T_p{}^q)_{n} {}^m {}_k {}^{l} \non \\
&=& \frac{1}{32\pi^2}
\left[ (\cW_\al \cW^\al)_{i}{}^{l} \de_k^j 
- (\cW_\al)_k{}^j (\cW^\al)_{i} {}^{l} 
- (\cW_\al)_{i} {}^{l} (\cW^{\al})_k {}^j 
+ (\cW_\al \cW^\al)_k{}^j \de_{i}^{l} \, \right]\,.
\eea
Similarly, the classical piece of 
$\bD^2 \bar{\hphi}_{\hi} {}^{\hj} (e^{\sf V} \hphi)_{\hk} {}^{\hl} $ 
is
$ -[\hW'(\hphi)_{\hi} {}^{\hj} - (\hb b)_{\hi} {}^{\hj} ] 
\hphi_{\hk}{}^{\hl} $
and the anomaly is 
\bea
&& \frac{1}{32\pi^2} (\hcW_\al)_{\hs}{}^{\hr} ( \hcW^\al)_{\hq}{}^{\hp} 
(T_{\hr}{}^{\hs})_{\hi}{}^{\hj} {}_{\hm} {}^{\hn} 
(T_{\hp}{}^{\hq})_{\hn} {}^{\hm} {}_{\hk} {}^{\hl} \non \\
&=& \frac{1}{32\pi^2}
\left[ (\hcW_\al \hcW^\al)_{\hi}{}^{\hl} \de_{\hk}^{\hj} 
- (\hcW_\al)_{\hk}{}^{\hj} (\hcW^\al)_{\hi} {}^{\hl} 
- (\hcW_\al)_{\hi} {}^{\hl} (\hcW^{\al})_{\hk} {}^{\hj} 
+ (\hcW_\al \hcW^\al)_{\hk}{}^{\hj} \de_{\hi}^{\hl} \, \right]
\eea

We will now consider various anomalous currents generalizing the 
above expressions. These currents all satisfy $0 = \vev{\bD^2 J}$
in any supersymmetric vacuum. 

\medskip
\noindent{\bf Quadratic relations}
\medskip

As a first example we consider the anomaly equation 
\bea \label{gU2quad1}
&& \!\!\!\! 0 = \ttpi \bvev{ \bD^2 \tr\!\left( \bar{\hb} \, 
e^{\sf V} \frac{\cW_\al}{z-\phi} \,b  \,\frac{\hcW^\al}{z-\hphi} \right) }
= \ttpi \bvev{ \left( \frac{\cW_\al}{z-\phi} 
\right)_{\raisebox{5pt}{\scriptsize $j$}}
{}^{\raisebox{2pt}{\scriptsize $k$}}  
\left( \frac{\hcW^\al}{z-\hphi} 
\right)_{\raisebox{5pt}{\scriptsize $\hl$}}
{}^{\raisebox{2pt}{\scriptsize $\hi$}}   
 \bD^2 
\bar{\hb}_{\hi}{}^j  
(e^{\sf V} b)_k{}^{\hl}  }
\non \\
&& \!\!\!\! = \ttpi \left\{\!
\bvev{ \! \tr \! \left(\frac{\cW_\al\cW^\al}{z-\phi}\right)
\tr \! \left(\frac{\hcW_\beta\hcW^\beta}{z-\hphi}\right) \! }
-  \bvev{ \! \tr \! \left( \frac{\phi}{z-\phi} b 
\frac{\hcW_\al\hcW^\al}{z-\hphi} \hb \right) \! }
+  \bvev{ \! \tr \! \left(\frac{\cW_\al\cW^\al}{z-\phi} 
b \frac{\hphi}{z-\hphi}\hb \right) \! } \!\right\}
\non \\
&&\!\!\!\! = R(z) \hR(z) + 
\ttpi \bvev{ \tr\!\left(b \frac{\hcW_\al \hcW^\al}{z-\hphi} \hb \right)} 
- \ttpi \bvev{ \tr\!\left( \hb \frac{\cW_\al \cW^\al}{z-\phi} b \right)} 
\eea
where we introduce
\be
R(z) \equiv -\frac{1}{32\pi^2} \bvev{ 
\tr \!\left(\frac{\cW_\al \cW^{\al} }{z-\phi} \right) } \,, \qquad
\hR(z) \equiv -\frac{1}{32\pi^2} \bvev{ 
\tr \!\left(\frac{\hcW_\al \hcW^{\al} }{z-\hphi} \right) }.
\ee
A few words of explanation are in order. We have 
dropped covariantization with $e^{\sf V}$ and $e^{-{\sf V}}$  since 
this will not affect the chiral part \cite{Cachazo:2002b}. We have used 
 (\ref{bhbanom}) together with the fact that in 
the chiral ring no more than two $\cW_\al$'s and $\hcW_\al$'s can 
have their gauge indices 
contracted \cite{Cachazo:2002b}. In addition we  
have also used the factorization of the expectation values in the 
chiral ring \cite{Novikov:1983,Cachazo:2002b}, and made repeated 
use of the relations (\ref{U2chirels}). Similar considerations will 
be used throughout this section.

Next we consider (here $f(z)$ is a polynomial; see \cite{Cachazo:2002b} 
for similar calculations)
\bea \label{U2first}
&& \!\!\!\!0 = \ttpi  \bvev{ \bD^2 \tr\! \left( \bar{\phi} \, e^{\sf V} 
\frac{f(\phi)\cW_\al \cW^\al}{z-\phi} \right)} 
 =  R(z)^2 f(z) - \ttpi \bvev{ \tr\!\left( \hb 
\frac{f(\phi)\cW_\al \cW^{\al}}{z-\phi} b \right) }  \\ 
&& \!\!\!\!- \frac{1}{(32\pi^2)^2} \sum_m c_m \sum_{k=0}^{m-1} 
\tr(\phi^k \cW_\al \cW^\al) \tr(\phi^{m-k-1}\cW_\beta \cW^\beta) 
 + \,\ttpi \bvev{ \tr\!\left( \frac{W'(\phi) 
f(\phi) \cW_\al \cW^{\al}}{z-\phi} \right) } \,, \non
\eea
where we have used (\ref{phianom}) together with the fact that 
$\frac{\f(z)-\f(\phi)}{z-\phi}  \equiv \sum_{m} c_m \phi^m$ 
is a polynomial.      

In particular, for $f(z) =1$ we get
\be \label{gU2quad2}
R(z)^2  - W'(z) R(z) 
= \frac{1}{32\pi^2} 
\bvev{ \tr\!\left( \hb \, \frac{\cW_\al \cW^\al}{z-\phi} \, b \right)}
+\frac{1}{32\pi^2} \bvev{ \tr\!\left( 
\frac{W'(z) - W'(\phi)}{z-\phi} \cW_\al \cW^\al \right)}. 
\ee

Analogously, one readily derives 
\bea \label{U2second}
&& \!\!\!\!0 = \ttpi  \bvev{ \bD^2 \tr\! \left( \bar{\hphi} \, e^{\sf V} 
\frac{f(\hphi)\hcW_\al \hcW^\al}{z-\hphi} \right)} 
 =  \hR(z)^2 f(z) + \ttpi \bvev{ \tr\!\left( b 
\frac{f(\hphi)\hcW_\al \hcW^{\al}}{z-\hphi} \hb \right) }  \\ 
&& \!\!\!\!- \frac{1}{(32\pi^2)^2} \sum_m c_m \sum_{k=0}^{m-1} 
\tr(\hphi^k \hcW_\al \hcW^\al) \tr(\hphi^{m-k-1}\hcW_\beta \hcW^\beta) 
 - \,\ttpi \bvev{ \tr\!\left( \frac{\hW'(\phi) 
f(\hphi) \hcW_\al \hcW^{\al}}{z-\hphi} \right) } \,, \non
\eea
and
\be \label{gU2quad3}
\hR(z)^2  + \hW'(z) \hR(z) 
= -  \frac{1}{32\pi^2} 
\bvev{ \tr\!\left(b \, \frac{\hcW_\al \hcW^\al}{z-\hphi} \, \hb\right) }
-\frac{1}{32\pi^2} \bvev{ \tr\!\left( 
\frac{\hW'(z) - \hW'(\hphi)}{z-\hphi} \hcW_\al \hcW^\al \right)}.
\ee

The similarity of eqs.~(\ref{gU2quad1}), (\ref{gU2quad2}) and (\ref{gU2quad3}) 
with eqs.~(\ref{homeom}), (\ref{exp2}) and (\ref{exp3}) is obvious.
Combining eqs. (\ref{gU2quad1})-(\ref{gU2quad3}) to eliminate 
$b$ and  $\hb$,  we find
\be \label{gU2r1eq}
R(z)^2 + \hR(z)^2 - R(z) \hR(z) - W'(z) R(z) + \hW'(z)\hR(z) 
= r_1(z)
\ee
with
\be \label{gU2r1}
r_1(z) =  
 \ttpi \bvev{
\tr\!\left(\frac{W'(z)-W'(\phi)}{z-\phi} \cW_\al\cW^\al \right) } 
- \ttpi \bvev{\tr\!\left(\frac{\hW'(z)-\hW'(\hphi)}{z-\hphi} 
\hcW_\al \hcW^\al \right) } 
\ee
The above two equations are the gauge theory analogs of the matrix-model 
results (\ref{U2r1eq}) and (\ref{U2r1}). Since the effect of $r_1(z)$ 
in (\ref{gU2r1eq}) is to eliminate the positive powers in 
the Laurent expansion of $- W'(z) R(z) + \hW'(z)\hR(z)$, the polynomial 
$r_1(z)$ has the same function as in the matrix model. The two equations are 
therefore equivalent and we may identify
\be \label{Rom}
R(z) = \om(z) \,, \qquad \hR(z) = \home(z)\,.
\ee
By looking at the other equations above one may also identify 
\bea
\gs \bvev{ \!\tr\!\left(B \, f(\hPhi)  \,\hB \right)\!} 
&=& -\ttpi \bvev{ \!\tr\left(b \, f(\hphi) \hcW_\al\hcW^\al \,\hb \right)\!} 
\,, \non \\
\gs \bvev{\! \tr\!\left(\hB \, f(\Phi)  \, B \right)\!} 
&=& -\ttpi \bvev{\! \tr\left(\hb \, f(\phi) \cW_\al\cW^\al \, b\right)\!}\,.
\eea

\medskip
\noindent{\bf Cubic relations}
\medskip

Strictly speaking, equation (\ref{gU2r1eq}) involves both resolvents 
 so we need 
one more relation before we can make the identifications (\ref{Rom}). 
Such a relation is obtained if we can show that the cubic loop equation 
(\ref{U2s1eq}), (\ref{U2s1Phi}) is also realized in the gauge theory. 
Given the close correspondence 
between the gauge theory and matrix model expressions noted above, the 
only thing we need to check is that (\ref{stepI}), (\ref{stepII}) 
and (\ref{stepIII}) are also realized in gauge theory consistent with the 
above identifications. If this is true then the cubic equation 
will follow in the same way as in the matrix model analysis.

The gauge-theory analog of (\ref{stepI}) is
\bea
\!\!\!\!0 &=& \ttpi \bvev{ \bD^2 \,\tr\!\left( 
\bar{\hphi} \, e^{\sf V} \, \hb \, b \, 
\frac{\hcW_\al\hcW^\al}{z-\hphi}\right) } \, = \, 
-\ttpi \hR(z) \bvev{ \tr\!\left( 
b \frac{\hcW_\al \hcW^\al}{z-\hphi} \hb \right) } \non \\
&& -\,\ttpi \bvev{ \tr\!\left( 
b  \frac{\hW'(\hphi)}{z-\hphi} \hcW_\al \hcW^\al \hb \right)} 
+\ttpi \bvev{ \tr\!\left( \hb b \frac{\hcW_\al \hcW^\al}{z-\hphi} \hb b 
\right)}\,.
\eea
Similarly, consideration of 
$\ttpi \vev{ \bD^2 \,\tr\!\left(\bar{\phi} \, e^{\sf V} \, b \, \hb \, 
\frac{\cW_\al\cW^\al}{z-\phi} \right)}$  
leads to the analog of (\ref{stepII}):
\be 
0 = -\ttpi R(z) \bvev{\tr\!\left(\hb \frac{\cW_\al \cW^\al}{z-\phi} b \right)}
+ \ttpi \bvev{ \tr\!\left(\hb \frac{W'(\phi)}{z-\phi}\cW_\al \cW^\al b 
\right)} 
- \ttpi \bvev{ \tr\!\left(b \hb \frac{\cW_\al \cW^\al}{z-\phi} b \hb\right) 
}\,.
\ee
Finally, the analog of (\ref{stepIII}) is
\bea 
0 &=& \ttpi \bvev{ \bD^2 \, \tr\!\left( \bar{b} \, 
e^{\sf V} \frac{\hcW_\al }{z-\hphi} \, \hb \,b \, \hb 
\,\frac{\cW^\al}{z-\phi} \right) } \,=\, -\ttpi 
\hR(z) \bvev{ \tr\!\left( \hb \frac{\cW_\al \cW^\al}{z-\phi} b \right)}  \\
&& \!\!\!\!\!\!\! -\,  
\ttpi R(z) \bvev{ \tr\!\left(b \frac{\hcW_\al \hcW^\al}{z-\hphi} \hb 
\right) } 
+ \ttpi\bvev{ \tr\!\left( \hb \, b   \frac{\hcW_\al \hcW^\al}{z-\hphi} 
\hb \, b \right)}
- \ttpi\bvev{ \tr\!\left( b \, \hb \frac{\cW_\al \cW^\al}{z-\phi}  
b\, \hb \right)}\,.  \non 
\eea
This completes the discussion of the equivalence of the matrix-model 
loop equations and gauge-theory anomaly equations.

It is worth noting that we did not have to use the entire chiral ring 
(which includes expressions with arbitrary many bifundamental fields) to 
derive the equations which determine $R(z)$, $\hR(z)$.

\medskip
\noindent{\bf Relation between gauge-theory and matrix-model expectation values}
\medskip

It is also of obvious interest to look for equations which determine  
\be
T(z) \equiv \bvev{ \tr\!\left(\frac{1}{z-\phi}\right) }\,, \qquad
\hT(z) \equiv \bvev{ \tr\!\left(\frac{1}{z-\hphi}\right)}\,,
\ee
since these expressions act as generating functions for the gauge-theory 
expectation values $\tr(\phi^k)$ and $\tr(\hphi^k)$,
whereas $R(z)$ and $\hR(z)$ (which by the above analysis are
equal to $\om(z)$ and  $\home(z)$, respectively)
are the generating functions (\ref{ress}) 
for the matrix-model expectation values $\tr(\Phi^k)$ and $\tr(\hPhi^k)$. 

Before discussing the $\U(N){\times}\U(N)$ case, let us recall 
the case of the $\U(N)$ theory with 
adjoint matter only. In a supersymmetric vacuum the equations 
governing this model are~\cite{Cachazo:2002b} 
\bea \label{Ukon}
&&R(z)^2 -W'(z) R(z) = \fourth f(z) 
= \ttpi \bvev{ \tr\!\left(\frac{W'(z)-W'(\phi)}{z-\phi}\cW_\al\cW^\al
\right)}\,,  \\
&&2 R(z) T(z) - W'(z) T(z) = \fourth c(z) 
= - \bvev{\tr\!\left(\frac{W'(z)-W'(\phi)}{z-\phi}\right) } \non\,.
\eea
Recalling the definition of the glueball field, 
$S = -\ttpi\tr(\cW_\al \cW^\al)$,
we see that the second equation is formally the 
derivative of the first equation, with the identifications 
$T(z) = \frac{\pa}{\pa S} R(z)$
and $c(z)  = \frac{\pa}{\pa S} f(z)$.
On the gauge theory side, this of course does not quite make sense; 
on the matrix model side, however,  where $S$ is just a parameter ($=\gs M$), 
it makes sense to take a derivative with respect to $S$. 
Since $R(z)$ in the gauge theory is identified with $\om(z)$ in the 
matrix model,
we are therefore led to the equation\footnote{\label{Sfoot}Here, 
and in subsequent equations, 
$\partial/\partial S$  should be identified with 
$ \sum_i N_i \partial/\partial S_i $,
as can be seen by considering the $z\to\infty$ part
of this equation. 
When we compare with the results in ref.~\cite{Naculich:2002b},
we set $N_i=1$ for all $i$.}
\be \label{adjT}
T(z) = \frac{\pa}{\pa S} \om\,.
\ee
Precisely this formula
was proposed in refs.~\cite{Gopakumar:2002,Naculich:2002b} (taking 
into account differences in conventions 
and recalling that $\om$ in the formula above 
is only the leading term in the genus expansion of the resolvent, $\oms$).

A similar analysis can be carried out in the $\U(N)$ model with additional 
matter in the fundamental representation and a superpotential of the form 
$\tr [ W(\phi)] + \sum_{I=1}^{N_f} \tilde{q}^I (\phi + m_I) q^I$.  
For this particular case we have, using the results in~\cite{Seiberg:2002}
\bea \label{Konfund}
&&\!\!\!\!\!\!\!\!\!\!\!\!R(z)^2 -W'(z) R(z) = \fourth f(z) 
= \ttpi \bvev{ \tr\!\left(\frac{W'(z) 
- W'(\phi)}{z-\phi}\cW_\al\cW^\al\right)} \,,
\non \\
&&\!\!\!\!\!\!\!\!\!\!\!\! 2 R(z) T(z) - W'(z) T(z) 
- \sum_{I=1}^{N_f} \tilde{q}^I \frac{1}{z-\phi} q^I 
= \fourth c(z) 
= - \bvev{\tr\!\left(\frac{W'(z)-W'(\phi)}{z-\phi}\right) }\,.
\eea
 The first equation is the same as in the case without fundamental 
matter (\ref{Ukon}), but the second equation has an extra contribution.
We also have \cite{Seiberg:2002} 
\be
\tilde{q}^I \frac{(\phi + m_I)}{z-\phi} q^I = R(z) \qquad 
\mbox{(no sum over $I$)}\,.
\ee
It follows from this equation that
\be
\frac{R(z)}{z+m_I} = \tilde{q}^I \frac{1}{z-\phi} q^I - 
\tilde{q}^I \frac{1}{z+m_I} q^I = 
\tilde{q}^I \frac{1}{z-\phi} q^I + \frac{R(-m_I)}{z+m_I} \,.
\ee
Using this result to eliminate the $q$-dependence 
in (\ref{Konfund}) and using the argumentation above we 
are led to the identification (where $R(z) = \oms(z)$)
\be \label{fundT}
T(z)=\frac{\pa }{\pa S}\oms(z) + \frac{1}{2\oms(z) 
- W'(z)}\sum_I \frac{\oms(z)-\oms(-m_I)}{z+m_I}
= \frac{\pa}{\pa S} \oms(z) + \omp(z)\,.
\ee
where $\omp(z)$ is the subleading (disk) contribution \cite{Naculich:2002b}
in the topological expansion of the resolvent:
$\om(z) = \oms(z) + \gs \omp(z) + \cdots$.
Note that the expression (\ref{fundT}) 
precisely agrees with eqs.~(8.7), (8.9) 
in (version 3 of) ref.~\cite{Naculich:2002b} 
(after taking into account differences in conventions:
$\oms= - S \om_{\rm s}$, $\omp = - \om_{\rm d}$). 
It is interesting to note that on the matrix-model side the extra term 
in~(\ref{fundT}) compared to (\ref{adjT}) comes from a subleading (disk) 
effect in the 
matrix-model loop equation \cite{Naculich:2002b}, whereas on the 
gauge theory side it 
arises from another equation, rather than from a subleading term. 
We also note that (\ref{fundT}) and the more explicit expression derived 
from it (eq.~(8.15) in \cite{Naculich:2002b}, valid when $N_f<N$) 
agrees with (a special case of) the  expression for 
$T(z)$ given in the very recent paper \cite{Cachazo:2003b} 
(cf.~eqs~(3.10), (3.11) of that paper),
using in particular the result 
$\vev{\tr(\phi^k)} = \vev{\tr(\phi^k)}_{\rm classical}$ 
for $k \le N$.

Let us now return to the $\U(N){\times}\U(N)$ model. By repeating 
the steps which lead to (\ref{gU2r1eq}) using analogous currents, but 
without the  $\cW_{\al}\cW^\al$ and $\hcW_\al \hcW^\al$ factors, 
one may derive 
\bea \label{gU2c1eq}
&& \!\!\!\! \! 0 =  
-\bD^2 \bvev{ \tr\left( \bar{\phi} \, e^{\sf V} \frac{1}{z-\phi} \right) }
- \bD^2 \bvev{ \tr\left(\bar{\hphi} \, e^{\sf V} \frac{1}{z-\hphi} \right) }
+ \bD^2 \bvev{ \tr\left( \bar{\hb} \, e^{\sf V} \frac{1}{z-\phi} 
                          \,b  \,\frac{1}{z-\hphi} \right) } \\
&& = \,  2R(z) T(z)  + 2\hR(z) \hT(z) 
- R(z) \hT(z) - \hR(z) T(z) 
 - W'(z) T(z) + \hW'(z) \hT(z) - c_1(z) \non 
\eea
(where we have dropped terms involving -- after factorization --
$\vev{\tr(\frac{\cW_\al}{z-\phi})}$ and 
$\vev{\tr(\frac{\hcW_\al}{z-\hphi})}$ since such terms vanish 
in a supersymmetric vacuum) and 
the polynomial $c_1(z)$ is explicitly given by 
\be \label{gU2c1}
c_1(z) = - \bvev{ \tr\left(\frac{W'(z)-W'(\phi)}{z-\phi} \right) } 
+  \bvev{\tr\left(\frac{\hW'(z)-\hW'(\hphi)}{z-\hphi} \right) } \,.
\ee
We note that (\ref{gU2c1eq}) can formally be obtained by taking 
derivatives of (\ref{gU2r1eq}).
We are therefore led to suggest the identifications\footnote{This
result may also be obtained
via the method in refs.~\cite{Naculich:2002a, Gopakumar:2002}.}
\be \label{U2Tom}
T(z) = \left[ \frac{\pa}{\pa S} + \frac{\pa}{\pa \hS} \right] \om\,, \qquad 
\hT(z) = \left[ \frac{\pa}{\pa S} + \frac{\pa}{\pa \hS} \right] \home \,.
\ee
which fit nicely into the structure given by the results (\ref{adjT}), 
(\ref{fundT}). 
Since we have two unknowns but only one equation, we can 
not argue unambiguously in favor of the above identifications, but 
consideration of the cubic equation for $T(z)$, $\hT(z)$ analogous 
to the one for $R(z)$, $\hR(z)$ presumably 
also leads to (\ref{U2Tom}), although we have not checked this explicitly.

\setcounter{equation}{0}
\section{{\large $\U(N)$} with symmetric or antisymmetric matter: I} 
\label{sUSAI}

In this section we consider the $\cN=1$ $\U(N)$ 
supersymmetric gauge theory with one chiral superfield 
$\phi_i{}^j$  transforming in the adjoint representation 
of the gauge group, 
one chiral superfield $x_{ij}$ transforming in either the symmetric  
($\Ysymm$)
or the antisymmetric 
($\Yasymm$) 
representation, and one chiral 
superfield $\tx^{ij}$
transforming in the conjugate representation. 
We treat the cases of the symmetric and 
antisymmetric representations simultaneously  
by assuming that $x$, $\tx$ satisfy 
$x^T = \beta x$ and $\tx^T = \beta \tx$, 
where $\beta =1$ for the symmetric representation, 
and $\beta =-1$ for the antisymmetric representation.

The superpotential of the gauge theory is taken to be of the 
form\footnote{We do not explicitly include a mass term for the 
$x$, $\tx$ fields although we think of these fields as being 
massive, cf.~footnote \ref{massfoot}.}
\be \label{USAsuppot}
\cW(\phi,x,\tx) = \tr[W(\phi) - \tx\, \phi \,x]\,, 
\ee
where $W(\phi) = \sum_{m=1}^{N+1} (g_m/m) \tr(\phi^m)$. 
This superpotential can be viewed as a deformation of an $\cN=2$ theory.
 
Below, after deriving the loop equations of the matrix model
(including the first subleading contribution in the $1/M$ expansion)
we establish the non-perturbative equivalence of the holomorphic sector of  
the above gauge theory to the associated matrix model,
by showing how the matrix-model loop equations are encoded 
in the gauge theory.  

The extension of the perturbative argument given 
in ref.~\cite{Dijkgraaf:2002e} to include the models 
considered in this section will be treated in section \ref{sUSAII}. 

\subsection{Matrix model analysis}

The partition function for the (holomorphic) matrix model is taken to 
be\footnote{As in the previous section, we use capital letters 
to denote matrix model quantities. All matrix indices run over $M$ values.}
\be \label{USAZ}
Z = \int \D \Phi \, \D X \, \D \tX \, 
\e^{ -\frac{1}{\gs}\tr[W(\Phi) - \tX \Phi X] } \,,
\ee
where $X^T = \beta X$, $\tX^T=\beta\tX$, and $\beta=1$ ($-1$) for  
$\Ysymm$ ($\Yasymm$).

We are interested in the planar limit of the matrix model, i.e.~the 
limit in which $\gs \rar 0$ and $M\rar \infty$, keeping 
$S=\gs M$ fixed.
The above matrix model is closely related to the $O(n)$ 
matrix model~\cite{Kostov:1989} with $n=1$. 
The planar saddle-point solution of that model 
was derived in refs.~\cite{Eynard:1992};
see also the recent paper \cite{Klemm:2003} 
where the planar solution to (\ref{USAZ}) 
was discussed. 
In the saddle-point approach, 
one diagonalizes the matrix $\Phi$ 
and derives equations satisfied by the resolvent\footnote{We use an
unconventional normalization of the resolvent in order to 
make the comparison with gauge theory more transparent.
Also, in order 
not to clutter the formul\ae{} we drop the $\vev{\cdots}$ when writing 
expressions in terms of eigenvalues.} 
\be \label{res}
\om(z) = \gs \bvev{ \tr\left(\frac{1}{z-\Phi}\right) } = 
\gs \sum_i \frac{1}{z-\la_i} \,,
\ee
where matrix-model expectation values are defined via
\be \label{USAMMexp}
\vev{ \cO(\Phi,X,\tX) } = 
\frac{1}{Z} \int \D \Phi \, \D X \, \D \tX \,\cO(\Phi,X,\tX)  \,
\e^{ -\frac{1}{\gs}\tr[W(\Phi) - \tX \Phi X] }\,.
\ee
Some details of the saddle-point analysis are given in appendix \ref{appcon}. 

Below we derive the equations satisfied by the resolvents
using an approach \cite{Cachazo:2002b}
(see also the approach in e.g. \cite{Kostov:1989, Eynard:1992})
that is close in spirit to the gauge theory analysis 
given in section \ref{sgUSA}. 
The discussion closely parallels the one in section \ref{sU2} 
(which is not surprising since the models in this section 
are orientifolds\footnote{Most of the equations 
in this section can be 
related to the ones in section \ref{sU2} by implementing an 
orientifold projection on the fields. Note, however, that the 
subleading terms to be discussed below 
can {\em not} be obtained this way.} of the one in 
section \ref{sU2}). 
(We stress that this method does not assume that the matrices are hermitian.)

\medskip
\noindent{\bf Quadratic relations}
\medskip

We begin by considering the Ward identity
\bea \label{USAquad1}
0 &=& \frac{2\gs^2}{Z} \int \D \Phi \, \D X \,  \D \tX  \, 
\frac{\D}{\D X_{ij}}  \left\{
\left[ \frac{1}{z-\Phi} X \left( \frac{1}{z+\Phi}\right)^T  
\right]_{\raisebox{5pt}{$\scriptstyle ij$}}
\, \e^{ -\frac{1}{\gs}\tr[W(\Phi) - \tX \Phi X] } \right\} \non \\
&=& \gs^2 \bvev{ \tr\!\left(\frac{1}{z-\Phi}\right) 
                 \tr\!\left(\frac{1}{z+\Phi}\right) } 
+ \beta \gs^2 \bvev{ \tr\!\left(\frac{1}{z-\Phi} \frac{1}{z+\Phi}\right) } 
\\
&&
+ \gs \bvev{ \tr\!\left[ \frac{1}{z-\Phi} X 
   \left(\frac{1}{z+\Phi} \right)^T \!\tX \Phi \right] } 
+ \gs \bvev{ \tr\!\left[ \frac{1}{z+\Phi} X 
  \left(\frac{1}{z-\Phi} \right)^T \!\tX \Phi \right] }  \non\\
&=&  -\om(z) \om(-z)  
  + \frac{\beta \gs}{2 z}[\om(z) -\om(-z)] 
+ \gs \bvev{ \tr\!\left( \tX \frac{1}{z-\Phi} X \right) } 
- \gs \bvev{ \tr\!\left( \tX \frac{1}{z+\Phi} X \right)  } \,,\non
\eea  
where we have used factorization of the expectation values 
in the planar limit. 
The corrections to factorization go like $\frac{1}{M^2}$ (or $\gs^2$). 
In the above expression, we have neglected the $\frac{1}{M^2}$ corrections, 
but have kept the $\frac{1}{M}$ (or $\gs$) subleading terms. Note that 
it is only the terms that are proportional to $\beta$ that are subleading 
(this feature is true in all equations in this section);  
the $\gs$-dependence in the last two terms in (\ref{USAquad1}) is related to 
our normalizations of $X$, $\tX$ and $\om(z)$, and does not mean that 
these terms are subleading.

In complete analogy with (\ref{exp2}) and (\ref{later1})
one may derive
\be \label{omom}
\om(z)^2  - W'(z) \, \om(z)  =
- \gs \bvev{\tr\!\left(\frac{W'(z) - W'(\Phi)}{z-\Phi}\right) }
- \gs \bvev{ \tr\!\left(\tX  \frac{1}{z-\Phi} X \right)}\,, 
\ee
as well as
\be\label{later3}
\gs \bvev{ \tr\!\left( \tX  \frac{\f(z)-\f(\Phi)}{z-\Phi} X \right)}
= - \gs^2 \bvev { \tr\!\left[ \frac{\D}{\D \Phi} 
\left(\frac{\f(z) - \f(\Phi)}{z-\Phi}\right) \right] } 
+ \gs \bvev{\tr\!\left(\frac{\f(z)-\f(\Phi)}{z-\Phi} W'(\Phi)\right) }.
\ee

Combining (\ref{USAquad1}) with (\ref{omom}) to eliminate the 
$X$-dependent terms, one finds
\be \label{USAr1eq}
\om(z)^2 + \om(-z)^2 + \om(z) \om(-z) - W'(z) \om(z) -  W'(-z)\om(-z) 
= r_1(z) + \frac{\beta \gs}{2 z}[\om(z) -\om(-z)], 
\ee
where 
\bea \label{USAr1}
r_1(z)&=&  -  \gs \bvev{\tr\!\left(\frac{W'(z)-W'(\Phi)}{z-\Phi}\right) } 
- \gs \bvev{\tr\!\left(\frac{ W'(-z) - W'(\Phi)}{-z-\Phi}\right) }  \non\\
      &=&  -  \gs \sum_i \frac{W'(z)-W'(\la_i)}{z-\la_i} 
- \gs \sum_i \frac{ W'(-z) - W'(\la_i)}{-z-\la_i} 
\eea
is a (manifestly even) polynomial of degree at most $N-1$.

The $\beta$-dependent terms on the r.h.s.~of 
eq.~(\ref{USAr1eq}) are subleading in $\gs$ compared to the 
rest of the terms.  
We may expand the resolvent in powers of $\gs$, i.e.~in a topological 
expansion~\cite{Ambjorn:1992} as   
$\om(z) = \sum_{\chi\le 2}\gs^{2-\chi}\om_{1-\chi/2}(z) = \oms(z) 
+ \gs \,\omp(z) + \cdots$. Here $\chi$ is the Euler characteristic,  
the leading term is the sphere ($\chi=2$) contribution, and the next term 
is an $\R \PP^2$ ($\chi=1$) contribution.
Using this expansion to  solve (\ref{USAr1eq}) order-by-order we find 
(in agreement with~\cite{Eynard:1992})
\be \label{om0}
\oms(z)^2 + \oms(-z)^2 + \oms(z) \oms(-z) - W'(z) \oms(z) 
-  W'(-z)\oms(-z) = r_1(z) \,,
\ee
and 
\bea \label{om1}
2\, \oms(z) \, \omp(z) + 2 \,\oms(-z) \, \omp(-z) + \oms(z)\, \omp(-z) 
+ \oms(-z) \, \omp(z) \non \\  - W'(z) \,\omp(z) -  W'(-z)\,\omp(-z) 
- \frac{\beta}{2z} [\oms(z) - \oms(-z)] = 0 \,.
\eea

\medskip
\noindent{\bf The cubic algebraic curve}
\medskip

As we now discuss (see also refs.~\cite{Eynard:1992,Klemm:2003}) there 
is a cubic algebraic curve underlying the model. 
The linear term in eq.~(\ref{om0}) can be eliminated by defining
\be\label{USAomrelate}
\oms(z) = \uu_1(z) + \om_r(z), \qquad \qquad \oms(-z) = \uu_3(z) 
+ \om_r(-z)\,,
\ee
with
\be \label{USAomdef}
\om_r(z) = \twothirds W'(z)- \third W'(-z)\,,
\ee
giving 
\be \label{USAr0eq}
\uu_1(z)^2 + \uu_3(z)^2 + \uu_1(z) \uu_3(z) 
= r_0(z) +  r_1(z)\,,
\ee
with
\bea\label{USAr0}
r_0  (z) 
&=& \om_r^2(z) + \om_r^2(-z) + \om_r(z) \om_r(-z) \non\\
&=& \third \left[ {W'}^{2}(z) + {W'}^{2}(-z) - W'(z)W'(-z) \right] \,,
\eea
a polynomial of degree $2N$.
Multiplying eq.~(\ref{USAr0eq}) by $\uu_1(z) - \uu_3(z)$, 
we find \cite{Eynard:1992}
\be
\uu_1(z)^3 - r(z) \uu_1(z) = \uu_3(z)^3 - r(z) \uu_3(z) \equiv s(z) \,,
\ee
so that $\uu_1(z)$ and $\uu_3(z)$ are both roots of the cubic equation
\be\label{USAcubic}
0 = u^3 - r(z) \,u - s(z)  = (u-\uu_1(z)) (u-\uu_2(z)) (u-\uu_3(z)) \,.
\ee
The absence of the quadratic term show that the third root
is $\uu_2(z) = - \uu_1(z) - \uu_3(z)$, and 
\be\label{sSAdef}
s(z) = \uu_1(z) \uu_2(z) \uu_3(z) 
= [ \oms(z) - \om_r(z) ] [ -\oms(z) -\oms(-z) + \om_r(z) + \om_r(-z)]
[\oms(-z) - \om_r(-z)]\,,
\ee
which we will show to be a (manifestly even) polynomial below.

Defining $s(z)  = s_0(z) + s_1(z)$ with 
\bea\label{s0SAdef}
s_0(z) &=&  \om_r(z)\, \om_r(-z) [\om_r(z) + \om_r(-z)]  \non\\
      &=& {\textstyle{1 \over 27}} 
       [- W'(z)+2W'(-z)][2W'(z) - W'(-z)] [W'(z) + W'(-z)]\,,
\eea
a polynomial of degree $3N$,
we can rewrite the cubic equation as
\bea
r_1(z) \,u +  s_1(z)
&=& u^3 - r_0(z) u - s_0(z)  \non\\
&=& (u+\om_r(z)) (u-\om_r(z)-\om_r(-z)) (u+\om_r(-z)) \,.
\eea
{}From eqs.~(\ref{sSAdef}) and (\ref{s0SAdef}) it follows that
\bea \label{USAs1eq}
s_1(z) &=& - \, \oms(z) \, \oms(-z) [\oms(z) + \oms(-z)] 
+\twothirds  [ W'(z) + W'(-z)] \,\oms(z)\, \oms(-z)   \non\\
&& +\,\om_r(-z)\,  [ \oms(z)^2 - W'(z) \oms(z) ] 
+ \om_r(z) \, [\oms(-z)^2 - W'(-z) \oms(-z)]\,.
\eea
We will show below that $s_1(z)$ is an even polynomial of degree at most 
$2N{-}1$.

\medskip
\noindent{\bf Cubic relations}
\medskip

Using Ward identities, we now show how to obtain the relation 
(\ref{USAs1eq}) with an explicit expression for $s_1(z)$. 
In complete analogy with eq.~(\ref{stepI}) and (\ref{stepII}) we have
\bea \label{USAstepII}
\!\!\!\! \!\!\!\! 
0 &=& \gs \, \om(z) \bvev{ \tr\left(\tX \frac{1}{z-\Phi}X \right)} 
- \gs \bvev{ \tr\!\left(\tX \frac{W'(\Phi)}{z-\Phi} X\right) } 
+ \gs \bvev{ \tr\!\left(X \tX \frac{1}{z-\Phi} \tX X\right)}. 
\eea
It can also be shown that
\bea \label{USAstepI}
0 &=& 2\frac{\gs^2}{Z} \int \D \Phi \, \D X \,  \D \tX  \, 
\frac{\D}{\D X_{ij}}  \left\{
\left[ \frac{1}{z-\Phi} X \tX X \left( \frac{1}{z+\Phi}\right)^T  
\right]_{\raisebox{5pt}{$\scriptstyle ij$}}
\, \e^{ -\frac{1}{\gs}\tr[W(\Phi) - \tX \Phi X] } \right\} \non \\
&=& 
\gs \,\om(z) \bvev{ \tr\!\left( \tX \frac{1}{z+\Phi} X \right)  } 
- \gs \,\om(-z) \bvev{ \tr\!\left( \tX \frac{1}{z-\Phi} X \right) } 
+ \gs \bvev{ \tr\!\left( X \tX \frac{1}{z-\Phi} X  \tX \right) }  \non\\
&&- \, \gs \bvev{ \tr\!\left( X \tX \frac{1}{z+\Phi} X \tX \right) } 
+ {\beta \gs^2 \over z} \left[
     \bvev{ \tr\!\left(\tX \frac{1}{z-\Phi} X \right)} +
            \bvev{\tr\!\left(\tX \frac{1}{z+\Phi} X \right) } \right] .
\eea  
Combining eqs.~(\ref{USAstepII}) and (\ref{USAstepI}),
and using eqs.~(\ref{USAquad1}) and (\ref{omom}), we find
\bea \label{USAintermed}
&&\!\!\!\!\!\!\!\!\!\! -\om(z)\om(-z) \left[\om(z) + \om(-z) \right] =
- \gs \bvev{\tr\!\left( \tX \frac{W'(\Phi)}{z-\Phi} X \right)}
+ \gs \bvev{ \tr\!\left(X \frac{W'(\Phi)}{z+\Phi}\tX\right) }   \\
&&\qquad\qquad\qquad \!\!\!  
+ \, {\beta \gs\over 2z} \left[ \om(z)^2 - \om(-z)^2 \right]
- {\beta \gs^2 \over z} \bvev{ \tr\!\left( \frac{W'(\Phi)}{z-\Phi} \right)
             + \tr\!\left(\frac{W'(\Phi)}{z+\Phi}\right) }\,.   \non
\eea
Only the first two (non $\beta$-dependent) terms 
on the r.h.s.~of this equation contribute to the leading-order 
(sphere) piece on the 
l.h.s., i.e.~to ~$-\oms(z)\oms(-z) \left[\oms(z) + \oms(-z) \right] $.
Considering only the leading terms in eq.~(\ref{USAintermed}) and using  
eqs.~(\ref{USAquad1})--(\ref{later3}), 
one obtains (by comparison with (\ref{USAs1eq})) an explicit expression 
for $s_1(z)$ in terms of the matrix model vevs 
$\vev{ \tr(\Phi^k)}$ 
\bea
s_1(z) &=& 
- \gs\,\om_r(-z) \lvev{\tr\left(\frac{W'(z)-W'(\Phi)}{z-\Phi}\right)}
 - \gs^2 \lvev { \tr \left[ \frac{\D}{\D \Phi}
\left(\frac{W'(z) - W'(\Phi)}{z-\Phi}\right) \right] }  \non\\
&&
+ \gs \lvev{\tr\left(\frac{W'(z)-W'(\Phi)}{z-\Phi} W'(\Phi)\right) }
+ (z \to -z) \,.
\eea
Using eq.~(\ref{eigenvaluetrick}), we can rewrite this
more explicitly, in terms of the eigenvalues $\la_i$ of $\Phi$:
\bea 
s_1(z) &=&  - \,\gs \,\om_r(-z) \sum_i 
\left[ \frac{ W'(z)  - W'(\la_i)}{z-\la_i} \right]
 -  2\gs^2 \sum_{i \neq j}\frac{1}{\la_i-\la_j} 
\left[ \frac{W'(z) - W'(\la_i)}{z-\la_i} \right] 
\non\\
& & +\,\gs \sum_{i} W'(\la_i)
\left[ \frac{W'(z) - W'(\la_i)}{z-\la_i} \right] 
+ (z \to -z)\,.
\eea
Finally, using the saddle point equations (\ref{USAsaddle}), 
we may rewrite this as 
\be \label{USAs1final}
\!s_1(z) =  - \gs \,\om_r(-z) \sum_i 
\left[\frac{ W'(z)  - W'(\la_i)}{z-\la_i}\right]
 -  \gs^2 \sum_{i,j} \frac{1}{\la_i+\la_j}
\left[\frac{W'(z)-W'(\la_i)}{z-\la_i} \right]  + (z\to -z)\,.
\ee
{}From this expression it is clear that $s_1(z)$ is an even polynomial 
of degree at most $2N{-}1$.
The result (\ref{USAs1final}) also appeared recently in ref.~\cite{Klemm:2003}, 
although using a different method of derivation. Next we turn to the 
gauge theory analysis.

\subsection{Gauge theory analysis} \label{sgUSA}

Below we show that the matrix-model loop equations discussed above can 
be obtained in the gauge theory from generalizations of 
the Konishi anomaly equations. 

As in section \ref{sU2}, it is sufficient to study the chiral part 
of the anomaly equations. In the chiral ring we have 
$0= [ \bar{Q}^{\dot{\al}},D_{\al\dot{\al}}F] = 
{\cW_\al} F $,  
where $\cW_\al$ is the gauge spinor superfield; more explicitly, 
$\cW_\al = {\cW}_{\al}^A T^A$, where $T^A$ are 
the representation matrices appropriate for the 
action of the gauge field on the field $F$. Using the explicit expressions 
for $T^A$ given in appendix \ref{apprep} one obtains the identities 
(valid in the chiral ring) 
\be \label{USAchirels}
{}[\cW_{\al},\phi]=0  
\,,\qquad \cW_{\al} x = -x (\cW_{\al})^T \,, \qquad 
\tx \cW_{\al} = -(\cW_{\al})^T \tx\,,
\ee
which will be repeatedly used below.

The basic building blocks that we will need are the 
(anomalous) currents 
$\bar{\tx}{}^{ij} (e^{\sf V} x)_{kl}$ and  
$\bar{\phi}_i{}^j (e^{\sf V}\phi)_k{}^l$. 
Using the superpotential (\ref{USAsuppot}),
one finds the classical piece of 
$ \bD^2 \bar{\tx}{}^{ij} (e^{\sf V} x)_{kl}$ to be 
$ -(\tx\phi)^{[ij)} x_{kl} = - \half[\tx \phi + \phi^T \tx]^{ij} x_{kl} $;
the anomalous contribution \cite{Konishi:1984} is
\bea
&& \frac{1}{32\pi^2} (\cW_\al)_n{}^m (\cW^\al)_q{}^p 
(T_m{}^n)_{rs}^{ij} (T_p{}^q)_{kl}^{sr} \non \\
&=& \frac{1}{32\pi^2}
\left[ 2(\cW_\al \cW^\al)_{[k} {}^{[j} \de_{l)}^{i)}  
+ 2(\cW_\al)_{[k}{}^{[j} (\cW^\al)_{l)} {}^{i)} \right] \,.
\eea
(The {\raisebox{2pt}{$\scriptstyle [\,\,\, )$}} notation is 
explained in appendix \ref{apprep}.) 
The classical piece of $\bD^2 \bar{\phi}_i {}^j (e^{\sf V} \phi)_k {}^l$ 
is given by 
$ W'(\phi)_i {}^j  \phi_{k}{}^{l} 
- (x \tx)_i {}^j \phi_{k}{}^{l} $
and the anomaly is the same as in eq.~(\ref{phianom}). 

As in section \ref{sU2}, we now generalize these 
currents\footnote{One can argue \cite{Cachazo:2002b} that there should 
be no chiral, perturbative corrections to the anomalies of the   
currents, but what about 
non-perturbative corrections? The $\SU(2N) + \sYasymm$ theory does have 
composite Pfaffian operators which might affect the discussion. 
However, since we are dealing with $\U(N)$ it seems 
that such operators should not be present.}. 
The approach is very similar to the one in section \ref{sU2} so we 
suppress the 
details.

\medskip
\noindent{\bf Quadratic relation}
\medskip

It can be shown that 
\bea \label{gUSAr1eq}
0 &=& \frac{1}{32\pi^2} \lvev{ \bD^2 \left[  
   \tr\!\left(\bar{\phi}\,
e^V \frac{\cW_\al \cW^\al}{z-\phi}\right) 
+  \tr\!\left(\bar{\phi} \,
e^V \frac{\cW_\al \cW^\al}{-z-\phi} \right) 
+ 2 \,\tr\!\left(\bar{\tx} \, e^{V} 
 \frac{\cW_\al}{z-\phi} x \left( \frac{\cW^\al}{z+\phi} \right)^T \right) 
\right] }
 \non \\
&=& R(z)^2 + R(-z)^2 + R(z) R(-z) - W'(z) R(z) - W'(-z) R(-z) - r_1(z)\,,
\eea
with
\be \label{gUSAr1}
r_1(z) = \ttpi \bvev{ 
\tr\!\left(\frac{W'(z)-W'(\phi)}{z-\phi} \cW_\al\cW^\al \right) } 
+ \ttpi \bvev{\tr\!\left(\frac{W'(-z)-W'(\phi)}{-z-\phi} 
\cW_\al \cW^\al \right) } \,,
\ee
where we have used (here and throughout) the fact that in 
the chiral ring no more than two $\cW_\al$'s can have their gauge indices 
contracted~\cite{Cachazo:2002b}, together with the factorization 
property~\cite{Novikov:1983,Cachazo:2002b} 
and also the relations (\ref{USAchirels}), valid in the chiral ring. 

\medskip
\noindent{\bf Cubic relation}
\medskip
 
One may also derive (\ref{USAs1eq}) in the gauge theory. To show this it is 
sufficient to obtain the gauge theory analogues  of (\ref{USAstepII}) 
and (\ref{USAstepI}). This is done by considering
\bea
\!\!\!\!\!\!\!\!0 &=& \ttpi \bvev{ \bD^2\, 
\tr\!\left( \bar{\phi} \, e^{V} \, x \, \tx \, 
\frac{\cW_\al\cW^\al}{z-\phi} \right) } 
\, = \, -\ttpi R(z) \bvev{ \tr\!\left( \tx \frac{\cW_\al \cW^\al}{z-\phi} x 
\right) }
 \non \\ &&+\, \ttpi \bvev{ \tr\!\left( \tx  \frac{W'(\phi)}{z-\phi} 
\cW_\al \cW^\al x \right) } 
-\ttpi \bvev{ \tr\!\left( x \tx  \frac{\cW_\al \cW^\al}{z-\phi} 
x \tx \right)}\,,
\eea
and
\bea 
&& \!\!\!\!\!\! 0 = -\ttpi \bvev{ \bD^2 \, \tr\!\left( \bar{\tx} \, 
e^{V} \frac{\cW_\al }{z-\phi} \, x \, \tx \, x 
\, \left( \frac{\cW^\al}{z+\phi} \right)^T \right)} 
= -\ttpi R(z) \bvev{ \tr\!\left( \tx 
\frac{\cW_\al \cW^\al}{z+\phi} x \right)} \non \\ 
&&\!\!\!\!\!\!+ \,\ttpi R(-z) 
\bvev{ \tr\!\left( \tx \frac{\cW_\al \cW^\al}{z-\phi} x \right) } 
- \ttpi \bvev{ \tr\!\left( x \, \tx   \frac{\cW_\al \cW^\al}{z-\phi} 
x \, \tx \right)}
+\ttpi \bvev{ \tr\!\left( x \, \tx \frac{\cW_\al \cW^\al}{z+\phi}  
x\, \tx \right) } \! \non 
\eea
By comparison of the leading ($\beta$-independent) parts of 
the matrix-model expressions with the above gauge theory equations, 
we find that they agree provided we identify 
\be
 R(z) = \om_0(z) \,.
\ee
We note that it was not necessary to consider the entire chiral ring 
(i.e.~operators with arbitrary many symmetric or antisymmetric fields) 
to obtain this result.
It is also possible to derive relations between expectation values 
involving the symmetric (or antisymmetric) fields, e.g.
\be
\qquad -\ttpi \bvev{ \tr(\tx \,f(\phi) \cW_\al \cW^\al \, x) } 
= \gs \bvev{ \tr(\tX \,f(\Phi)\, X) }\,.
\ee
Notice that no subleading terms appeared in the gauge theory equations. 
The role of the subleading terms in the matrix model expressions 
will become clear below.

\medskip
\noindent{\bf Relation between gauge-theory and matrix-model expectation values}
\medskip

The generating function for the gauge theory expectation values 
$\vev{\tr(\phi^k)}$ is
\be
T(z) \equiv \bvev{\tr\!\left( \frac{1}{z-\phi} \right)}\,.
\ee
An equation involving this function, analogous to (\ref{gUSAr1eq}), 
can be derived by dropping the $\cW_\al\cW^\al$ factors in the above 
currents, i.e.
\bea \label{gUSAc1eq}
&& \!\!\!\!0 =  
-\bD^2  \bvev{ \tr\!\left(\bar{\phi}\, e^V \frac{1}{z-\phi} \right) }
-\bD^2 \bvev{ \tr\!\left(\bar{\phi} \,e^V \frac{1}{-z-\phi} \right) } 
+ 2\bD^2 \bvev{ \tr\!\left( \bar{\tx} \,e^{V} 
 \frac{1}{z-\phi} x \left( \frac{1}{z+\phi} \right)^T \right) }
 \non \\
&& \!\!\!\!=  2\, R(z) T(z)  + 2\, R(-z) T(-z) 
+ R(z) T(-z) + R(-z) T(z)  \non \\ 
&& \!\!\!\! \quad- \,W'(z) T(z) - W'(-z) T(-z) - c_1(z) 
- \frac{2\beta}{z}[R(z)-R(-z)] \,,
\eea
where the polynomial $c_1(z)$ is explicitly given by 
\be
c_1(z) = - \bvev{ \tr\!\left(\frac{W'(z)-W'(\phi)}{z-\phi} \right) } 
-  \bvev{\tr\!\left(\frac{W'(-z)-W'(\phi)}{-z-\phi} \right) } \,.
\ee
As in section \ref{sgU2}, we note that, were it not for the $\beta$-dependent 
terms,  eq.~(\ref{gUSAc1eq}) could be viewed as the formal derivative 
of eq.~(\ref{gUSAr1eq}),
provided that  $c_1 =  \frac{\pa}{\pa S}r_1$;  
$T = \frac{\pa}{\pa S}R = \frac{\pa}{\pa S}\om_0$.  
To deal with the $\beta$-dependent 
terms we recall eq.~(\ref{om1}) and note 
that the identification\footnote{Here
$\partial/\partial S = \sum_i N_i \partial/\partial S_i $.
See footnote \ref{Sfoot}.}
\be \label{Tom}
T(z) = \frac{\pa}{\pa S} \oms + 4 \omp 
\ee
resolves the discrepancy.
Our suggested  expression (\ref{Tom}) generalizes the formula proposed 
in ref.~\cite{Naculich:2002b} (see 
also~\cite{Gopakumar:2002,Naculich:2002a,Cachazo:2002b}). 
Consideration 
of cubic equations involving $T(z)$ and $R(z)$ presumably also 
leads to (\ref{Tom}).

\setcounter{equation}{0}
\section{{\large $\U(N)$} with symmetric or antisymmetric matter: II} 
\label{sUSAII}

In this section we discuss how to extend the approach 
in ref.~\cite{Dijkgraaf:2002e} to the case with matter in the symmetric and 
antisymmetric representations. We will first give a heuristic argument 
and then give a more detailed argument and present an explicit 
sample calculation. 

Going through steps similar to the ones carried out 
in \cite{Dijkgraaf:2002e}, it can be shown that,
for the purpose of determining the effective action,
the superspace action for a field $\vphi_R$ in 
some non-real representation $R$ of $\U(N)$, together a field 
$\tilde{\vphi}_{\bar{R}}$ in 
the representation conjugate to $R$,  can be rewritten as 
\be
\int \D^4 x \,\D^2\theta \left[ -\frac{1}{2} \tilde{\vphi}_{\bar{R}} ( \Yfund 
+ m - i \cW_\al D_\al) \vphi_R + W_{\rm tree} (\vphi,\tilde{\vphi}) \right].
\ee

Following \cite{Dijkgraaf:2002e} we 
transform to momentum superspace ($p_\mu$, $\pi_\al$), where $\pi_\al$ 
is the fermionic momentum conjugate to the superspace coordinate 
$\theta^\al$. We write the propagator 
of the $n$th edge of a Feynman diagram in a Schwinger parameterization 
as \cite{Dijkgraaf:2002e}
\be
 \int_0^{\infty} \D s_n \e^{-s_n (p_n^2 + \cW^\al \pi_{n\al} + m)}\,.
\ee

Now, in standard double-line notation, the only difference (for planar 
diagrams drawn on the sphere) between the symmetric  (or antisymmetric)
representation and the adjoint one is that 
the orientation of one of the lines has changed\footnote{In addition 
the propagator for fields in the symmetric (or antisymmetric) 
representation has a 
twisted part not present for adjoint fields. This feature leads to 
the presence of planar diagrams drawn on $\sR \sPP^2$ in the 
topological expansion.}. 
Compared to the case with adjoint fields only, this means that 
each insertion of a $\cW_\al$ on a line with flipped orientation comes 
with a minus sign. However, since $W_{\rm eff}$ is a function of 
the glueball fields\footnote{For simplicity we restrict to the case of 
only one glueball field, $S=-\ttpi\tr(\cW_{\al} \cW^\al)$, in our 
explicit calculations, but 
whenever possible we write the formul\ae{} in their general form.}
$S_i = -\ttpi\tr(\cW_{i\al} \cW_i^\al)$ there are necessarily an even number 
(zero or two) of insertions on each line (index loop) and thus the extra 
minus signs cancel out and as in \cite{Dijkgraaf:2002e,Ita:2002} one finds 
\be 
W_{\rm eff} = \sum_i N_i \frac{\pa}{\pa S_i} F_{S^2}
+ 4  \, F_{\sR \sPP^2} \,.
\ee
Here the second piece arises from the twisted part of the propagator 
(equivalently, from planar diagrams on $\R \PP^2$). 
A gauge theoretic argument for the presence of this piece can  
be given along the lines of refs.~\cite{Ita:2002}. 
The factor of $4$ has the same origin as in the $\SO/\Sp$ models 
discussed in \cite{Ita:2002}. Also note that the factor of $4$ that appeared 
in (\ref{Tom}) is the same as the one in the equation above, as can been 
seen by using the methods in ref.~\cite{Naculich:2002a,Naculich:2002b}. 

On the other hand, the gauge-coupling matrix 
$\tau_{ij}(S)$ couples to $\tr(\cW^i_\al)\tr(\cW^{j\al})$ 
in the effective action and since insertions of a single $\cW_\al$ 
on an index loop coming from one of the lines with flipped orientation  
leads to a sign change, this implies that $\tau_{ij}$ will in general no 
longer will be 
given by $\frac{\pa^2 F}{\pa S_i \pa S_j}$\footnote{A similar discrepancy 
was observed in \cite{Balasubramanian:2002} in the context of 
multi-trace operators, but in that case the discrepancy appeared already 
in $W_{\rm eff}$.}. 
This is also clear from the point of 
view in \cite{Cachazo:2002b} where it was argued that the relation 
between $\tau_{ij}$ and $\frac{\pa^2 F}{\pa S_i \pa S_j}$ follows 
from a shift symmetry of the $\U(1)$ part of the gauge 
superfield $\cW_\al$. In the case of adjoint fields only, 
this symmetry  is a consequence of the decoupling of the
$\U(1)$ (since the adjoint action is via commutators). 
However, the symmetric (or antisymmetric) representation
couples to the $\U(1)$ (the gauge action is longer via commutators) so 
there is no shift symmetry and hence no direct relation 
between $\tau_{ij}$ and $\frac{\pa^2 F}{\pa S_i \pa S_j}$. 

However, 
even though $\tau_{ij}$ is not given by 
$\frac{\pa^2 F}{\pa S_i \pa S_j}$,
there is a simple way to keep track of the 
extra signs in matrix-model perturbation theory, i.e.~to determine 
$\tau_{ij}$ perturbatively from the  matrix model. 
To demonstrate this, we represent the $\tX \, \Phi \, X$ matrix-model 
vertex graphically as in figure 1.
\bigskip 
\begin{figure}[h]
\begin{center}
 \includegraphics{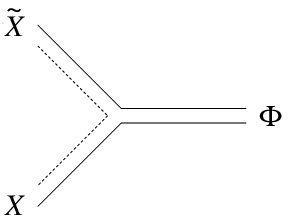}  \\[-0.1cm]
{\bf \caption{ \small \rm The $\tX \, \Phi \, X$-vertex. The dashed 
line has the opposite orientation compared to the $\Phi^3$ vertex.}}
\end{center}
\end{figure}
Now if one proceeds to calculate the matrix-model Feynman diagrams as 
usual, but for each index-loop constructed from a dashed line one makes 
the identification $g_{\rm s} M_i = \tilde{S_i}$, one will obtain 
a free energy of the form $F(S,\tilde{S})$. 
If one then takes the 
second derivative $\frac{\pa^2 F(S,\tilde{S})}{\pa S_i \pa S_j}$ 
using the rule 
$\frac{\pa \tilde{S}_i}{\pa S_j} = -\de_i^j$, 
and then afterwards sets $\tilde{S}_i = S_i$, 
one will obtain the right result for $\tau_{ij}$.
That is, the 
extra signs will be taken care of and the resulting $\tau_{ij}$ will 
agree with the gauge theory result.

We will now give more details for a specific set of diagrams 
(for simplicity we consider the case of a single glueball field, $S$). 
We consider the three gauge theory diagrams in figure~2. 
\bigskip \medskip
\begin{figure}[h]
\begin{center}
 \includegraphics{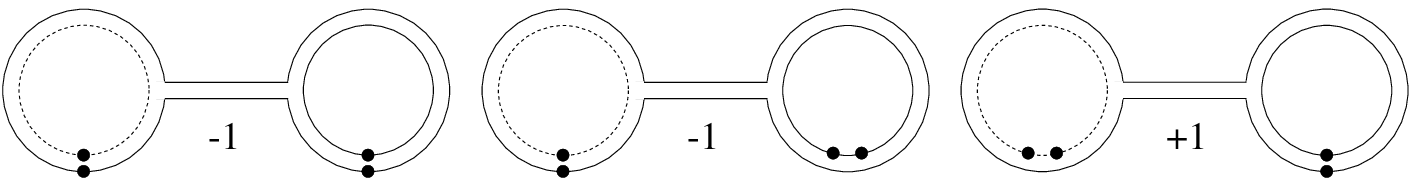}  \\[-3.0cm]
{\bf \caption{ \small \rm The black dots indicate $\cW_\al$ insertions. }}
\end{center}
\end{figure}
  
There are two Schwinger parameters corresponding to the momentum running 
in the two loops. Calling these $s_1$, $s_2$, 
the integral over bosonic 
momenta gives ${\rm const}\times(s_1 \, s_2)^{-2}$. 
The integral over fermionic 
momenta gives for the sum of the above three diagrams (plus an 
additional $\R \PP^2$ diagram not shown in the figure above) a 
constant times 
\bea \label{anti}
&& \!\!\!\!\!\!\!\! (s_1\,s_2)^2 (\cW_\al)_j{}^i (\cW^\al)_l{}^k 
(\cW_\beta)_n{}^m (\cW^\beta)_q{}^p
(T_i{}^j)_r{}^s{}_t{}^u (T_k{}^l)_u{}^t{}_s{}^v 
(T_m{}^n)_{va}^{bc} (T_p{}^q)_{cb}^{ar} 
\non \\ && \!\!\!\!\!\!\!\!= (s_1 s_2)^2 [
3N \tr(\cW_\al\cW^\al) \tr(\cW_\beta \cW^\beta) 
+ 4 \beta \tr(\cW_\al\cW^\al) \tr(\cW_\beta \cW^\beta)
+ 2 \tr(\cW_\beta \cW^\beta) \tr(\cW_\al) \tr(\cW^\al) ]
\non \\ &&
\!\!\!\!\!\!\!\! 
\propto (s_1 s_2)^2 [ N (3 S^2) + 4 \beta S^2 - \,S\, w_\al w^\al ]
\eea
where we have used the formul\ae{} in appendix \ref{apprep} as well as 
the definition $w_\al = \frac{1}{4\pi}\tr(W_\al)$. 
We see that the $s_i$ dependence cancels between the bosonic 
and fermionic momentum integrals as required for the reduction 
to a matrix model. 
For comparison, if all the fields had been in the adjoint representation, 
one would have obtained instead the same constant as above times 
\bea \label{adj}
&& (s_1\,s_2)^2 (\cW_\al)_j{}^i (\cW^\al)_l{}^k 
(\cW_\beta)_n{}^m (\cW^\beta)_q{}^p
(T_i{}^j)_r{}^s{}_t{}^u (T_k{}^l)_u{}^t{}_s{}^v 
(T_m{}^n)_v{}^a{}_b{}^c (T_p{}^q)_c{}^b{}_a{}^r 
\non \\ && = (s_1 s_2)^2 [
3N \tr(\cW_\al\cW^\al) \tr(\cW_\beta \cW^\beta) 
- 6 \tr(\cW_\beta \cW^\beta) \tr(\cW_\al) \tr(\cW^\al) ]
\non \\
&& \propto (s_1 s_2)^2 [ N (3 S^2) + 3\, S\, w_\al w^\al ]
\eea
By comparing (\ref{anti}) and (\ref{adj}) we see that the terms 
proportional to $N$ agree. These are contributions to 
$N\frac{\pa}{\pa S} F_{S^2}$. 
The second term in (\ref{anti}) contributes to $4F_{\sR \sPP^2}$ 
and comes from a diagram (not displayed in the figure above) with 
a twisted propagator. Finally, the last set of terms contribute to 
$\half \tau(s) w_\al w^\al$ and explicitly illustrate 
the sign rule discussed above. In the first case we have $-1-1+1 = -1$, and 
in the second case we get $+1+1+1 =+3$. The relative factor of 
$-\frac{1}{3}$ is indeed present in (\ref{anti}) vs. (\ref{adj}).

\setcounter{equation}{0}
\section{Summary} \label{sSum}

In this paper we focused on three $\cN=1$ supersymmetric 
gauge theories:
$\U(N){\times}\U(N)$ with matter in adjoint and bifundamental representations,
$\U(N)$ with matter in adjoint and symmetric representations, and 
$\U(N)$ with matter in adjoint and antisymmetric representations. 
As was shown, each of these theories exhibits a cubic algebraic curve. 
The equivalence of the matrix model and the gauge theory 
descriptions was established 
by means of generalized Konishi anomalies equations, which were 
shown to be equivalent to the matrix model loop equations. This 
result demonstrates the 
equivalence of the matrix models to the holomorphic sector of the 
gauge theories. 

In addition, we studied the relation between the generating functions
$T(z)$ of gauge theory vevs and the generating functions $\om(z)$
of matrix model vevs for each of the theories considered, 
generalizing the results of refs.~\cite{Gopakumar:2002,Naculich:2002b}.

We also 
investigated the matrix model/gauge theory equivalence 
using a perturbative superspace 
analysis. 
If matter in the symmetric (or antisymmetric) representation 
is present it was shown that the 
gauge-coupling matrix $\tau_{ij}$ is {\em not} given by the 
second derivative of 
the matrix model free energy; 
the latter must be modified diagram-by-diagram 
by suitably chosen minus signs in the matrix-model perturbative 
expansion so as to yield the correct gauge coupling matrix. 
As a result there does not appear at this point 
to be a concise formula expressing 
$\tau_{ij}$ in terms of the matrix model free energy, contrary to 
situations involving only adjoint, fundamental, or bifundamental matter.

\section*{Acknowledgments}
HJS would like to thank the string theory group and Physics 
Department of Harvard University for their hospitality extended 
over a long period of time.

\appendix
\section*{Appendices}

\setcounter{equation}{0}
\section{Saddle-point analysis} \label{appcon}
Here we will briefly discuss the saddle-point approach to study 
the planar solution of the matrix models discussed in this paper. In the 
recent papers \cite{Lazaroiu:2003,Klemm:2003} this approach has been 
extended to holomorphic matrix models.

\subsection{$\U(N){\times}\U(N)$ with bifundamental matter}  

The saddle-point approach to the $\U(N){\times}\U(N)$ quiver 
model has previously been discussed 
in refs.~\cite{Kostov:1992,Kharchev:1993,Dijkgraaf:2002b,Hofman:2002} 
(and was recently extended to holomorphic matrices in \cite{Lazaroiu:2003}). 
The first step is to transform to an eigenvalue basis 
for the adjoint fields and then integrate out the bifundamental fields. 
This reduces (\ref{U2Z}) to 
\be
Z \propto \int \prod_i \D \la_i\, \D \hla_i 
\frac{\prod_{i<j} (\la_i-\la_j)^2(\hla_i - \hla_j)^2}
{\prod_{i,j}(\la_i-\hla_j)}
\e^{-\frac{1}{g_s}\sum_i  \left[ W(\la_i)-\hW(\hla_i)\right] }\,.
\ee
The saddle-point equations of motion are thus
\bea\label{U2saddle}
-\frac{W'(\la_i)}{g_{s}} + 2 \sum _{j\neq i}\frac{1}{\la_i- \la_j}
-\sum_j \frac{1}{\la_i - \hla_j} &=& 0\,,\non \\
\frac{\hW'(\hla_i)}{g_{s}} + 2 \sum _{j\neq i}\frac{1}{\hla_i- \hla_j}
-\sum_j \frac{1}{\hla_i - \la_j} &=& 0\,.
\eea 
The equations (\ref{U2r1eq}) and (\ref{U2s1eq}) can be derived 
directly from the saddle-point 
equations. For instance, (\ref{U2r1eq}) can be obtained by considering 
\bea
&& \gs^2 \sum_i \frac{1}{z-\la_i}\left[ -\frac{W'(\la_i)}{g_{s}} 
+ 2 \sum _{j\neq i}\frac{1}{\la_i- \la_j} 
-\sum_j \frac{1}{\la_i - \hla_j} \right ] \non \\
&+& \gs^2  \sum_i \frac{1}{z-\hla_i}
\left[ \frac{\hW'(\hla_i)}{g_{s}} + 2 \sum _{j\neq i}\frac{1}{\hla_i- \hla_j}
-\sum_j \frac{1}{\hla_i - \la_j} \right] =0 \,.
\eea 
Alternatively, an expedient way to obtain 
(\ref{U2r1eq}) and (\ref{U2s1eq})  is by imposing 
\cite{Kharchev:1993,Dijkgraaf:2002b,Hofman:2002}
\be
\oint \D z \frac{1}{x-z} W^{(s)}(z) = 0\,,
\ee
where the contour encloses all eigenvalues but not the point $x$ and
the W-algebra currents $W^{(s)}(z)$ ($s=2,3$) are given by
\be
W^{(s)} = \frac{(-1)^s}{s} \sum_{i=1}^3 (\uu_i)^s\,,
\ee
where the $\uu_i$'s were defined in (\ref{U2omrelate}) and 
below (\ref{U2cubic}). 

\subsection{$\U(N)$ with (anti)symmetric matter }  

The first step of the saddle-point approach is to 
transform (\ref{USAZ}) into an eigenvalue basis 
for $\Phi$ and then integrate out $x_{ij}$ and $\tx^{ij}$. 
This leads to 
\be \label{USAeigen}
Z \propto \int \prod_i \D \la_i\, 
\frac{ \prod_{i<j} (\la_i-\la_j)^2 
\prod_i \la_i^{-\beta/2} }{\prod_{i,j}(\la_i + \la_j)^{1/2} }
\e^{-\frac{1}{g_s} \sum_i W(\la_i) }\,.
\ee
The saddle-point equation of motion is thus
\be \label{USAsaddle}
-\frac{W'(\la_i)}{g_{s}} + 2 \sum _{j\neq i}\frac{1}{\la_i- \la_j}
-\sum_j \frac{1}{\la_i + \la_j} -\frac{\beta}{2}  \frac{1}{\la_i}= 0\,.
\ee 
where the last term is a $1/M$ (or $\gs$) effect.

The model (\ref{USAeigen}) is closely related to the 
$O(n)$ matrix model~\cite{Kostov:1989} with $n=1$. 
The planar solution of that model was derived 
in~\cite{Eynard:1992}. 
(The extension to holomorphic matrices was recently 
discussed in ref.~\cite{Klemm:2003}.)

The expressions (\ref{USAr1eq}), (\ref{USAs1eq}) can be derived 
directly from the saddle-point equations. For instance, (\ref{USAr1eq}) 
can be obtained by considering 
\be
\sum_i \frac{1}{z-\la_i}\left[ -\frac{W'(\la_i)}{g_{s}} 
+ 2 \sum _{j\neq i}\frac{1}{\la_i- \la_j} 
-\sum_j \frac{1}{\la_i + \la_j}  -\frac{\beta}{2}  \frac{1}{\la_i} \right] 
+ (z\leftrightarrow -z) = 0\,.
\ee

Alternatively, an expedient way 
to obtain (\ref{USAr1eq}) and (\ref{USAs1eq}) is by imposing
\be
\oint \D z \frac{1}{x-z} W^{(s)}(z) = 0\,,
\ee
where the contour encloses all eigenvalues but not the point $x$ and
 the W-algebra currents $W^{(s)}(z)$ ($s=2,3$) are given by
\be
W^{(s)} = \frac{(-1)^s}{s} \sum_{i=1}^3 ({u}_i)^s\,,
\ee
where the $u_i$'s were defined in (\ref{USAomrelate}) and 
below (\ref{USAcubic}).

\setcounter{equation}{0}
\section{Some representation theory} \label{apprep}
Here we collect some explicit formul\ae{} for the 
generators in the various representations discussed in the main text.

\medskip
\noindent{\bf Adjoint representation of $\U(N)$ }
\medskip

In standard double-index notation the generators in the adjoint representation
are 
\be
(T_i{}^j)_k{}^{l} {}_{m} {}^{n} = \de_i^n \de_k^j \de_{m}^{l} 
- \de_i^{l} \de^j_{m} \de_k^n \,.
\ee
This gives the well-known results
\be 
(V \phi)_k{}^{l}
= V_j{}^i (T_i{}^j)_k{}^{l}{}_{m}{}^{n} \phi_n{}^{m} 
= [V,\phi]_k{}^{l}\,; \quad ({\cW_\al} \phi)_k{}^{l}
= (\cW_\al)_j{}^i (T_i{}^j)_k{}^{l}{}_{m}{}^{n} \phi_n{}^{m} 
= [\cW_\al,\phi]_k{}^{l}
\ee
where $V$ and $\cW_\al$ are the vector and spinor gauge 
superfields, respectively. 

\medskip
\noindent{\bf Bifundamental representations of $\U(N){\times}\U(N)$ }
\medskip

To describe the action of the gauge superfields on the bifundamental field 
$b_i{}^{\hj}$ it is convenient to use a composite index $I=(i,\hi)$. In 
this notation the gauge vector superfield 
is ${\sf V} =  V_J{}^I (T_I{}^J)$, where
\be
(T_I{}^J)_i{}^{\hj} {}_{\hk} {}^{l} = \de_I^l \de_i^J \de_{\hk}^{\hj} 
- \de_I^{\hj} \de^J_{\hk} \de_i^l \,,
\ee
and we have used the double-index notation. This implies 
\be 
({\sf V} b)_i{}^{\hj}
= V_J{}^I (T_I{}^J)_i{}^{\hj} {}_{\hk} {}^{l} b_l{}^{\hk} 
= V_i{}^k b_k{}^{\hj} - b_i{}^{\hk} \hV_{\hk} {}^{\hj} \,,
\ee
where $V$, $\hV$ are the gauge 
superfields for the two $\U(N)$ factors. 
One may view ${\sf V}$ as a diagonal $2{\times}2$ matrix,  
$\diag(V,\hV)$. In this notation $b$ and $\hb$ can 
be combined into an off-diagonal $2{\times}2$ matrix, and $\phi$ and $\hphi$ 
can be combined into a diagonal $2{\times}2$ matrix. In the $2{\times}2$ 
matrix notation, the gauge action is via commutators. 
Similarly the action of the gauge spinor superfield can be written as
\be
({\sf W}_\al b)_i{}^{\hj}
= (\cW_\al)_i{}^k b_k{}^{\hj} - b_i{}^{\hk} (\hcW_\al)_{\hk} {}^{\hj}\,,
\ee
where $\cW_\al$ and $\hcW_\al$ are the gauge spinor superfields  
corresponding to the two $\U(N)$ factors.

The action on the bifundamental field $\hb_{\hi}{}^j$ is the same as the 
one on $b$, but with tilde and un-tilde indices interchanged.

\medskip
\noindent{\bf  Symmetric/antisymmetric representation of $\U(N)$}
\medskip

The generators in the symmetric or antisymmetric representation 
of $\U(N)$ are 
\be
(T_i{}^j)_{kl}^{mn} = 2 \de_{[k}^j \de_{l)}^{[m} \de_i^{n)}\,,
\ee
where we have used the notation 
$u_{[i} v_{ j)} = \half (u_i v_j + \beta \, u_j v_i)$, where $\beta =+1$ 
for the symmetric representation and $\beta = -1$ for the antisymmetric 
representation. The action of $\cW_{\al}$ on $x$ is 
\be
({\cW}_\al x)_{kl}
= (\cW_{\al})_j{}^i (T_i{}^j)_{kl}^{mn} x_{nm} 
= [(\cW_\al)_k{}^j x_{jl} + \beta (\cW_\al)_{l}{}^{j} x_{jk}]\,,
\ee
or in matrix notation: $\cW_{\al} x + x (\cW_{\al})^T $.

The generators of semi-simple Lie algebras satisfy 
$\tr_R (T^A T^B) = I(R) \de^{AB}$ where
$I(R)$ is the index of the representation, i.e. $2N$ for the adjoint, $N-2$ 
for the antisymmetric and $N+2$ for the symmetric representation. 
 Using the above forms of the generators the expression $\tr_R (T^A T^B)$ will 
 contain some extra trace factors since we are dealing with $\U(N)$'s.

\providecommand{\href}[2]{#2}\begingroup\raggedright\endgroup

\end{document}